%% file: main.tex
\documentclass[conference]{IEEEtran}
\IEEEoverridecommandlockouts
\usepackage{cite}
\usepackage{amsmath,amssymb,amsfonts}
\usepackage{algorithmic}
\usepackage{graphicx}
\usepackage{textcomp}
\usepackage{algorithm}
\usepackage{subfigure}
\usepackage{multirow}
\usepackage{makecell}
\usepackage{diagbox}
\usepackage{amsthm}
\usepackage{mathtools}
\usepackage{xcolor}
\newtheorem{myDef}{DEFINITION}[section]
\newtheorem{myTheorem}{THEOREM}[section]

\newcommand{\nosection}[1]{\smallskip\noindent\textbf{#1.}}
\usepackage{enumitem}
\def\BibTeX{{\rm B\kern-.05em{\sc i\kern-.025em b}\kern-.08em
    T\kern-.1667em\lower.7ex\hbox{E}\kern-.125emX}}
\begin{document}

\title{DCMT: A Direct Entire-Space Causal Multi-Task Framework for Post-Click Conversion Estimation}

\author{
\IEEEauthorblockN{Feng Zhu{$^{1}$}, Mingjie Zhong{$^{1}$}, Xinxing Yang{$^{1}$}, Longfei Li{$^{1}$}, Lu Yu{$^{1}$}, Tiehua Zhang{$^{1}$}, Jun Zhou{$^{1*}$},\\ Chaochao Chen{$^{2}$}, Fei Wu{$^{2}$}, Guanfeng Liu{$^{3}$}, and Yan Wang{$^{3}$}} \\
\IEEEauthorblockA{\textit{{$^{1}$}Ant Group, Hangzhou, China} \\ 
\textit{{$^{2}$}College of Computer Science and Technology, Zhejiang University, Hangzhou, China} \\ \textit{{$^{3}$} School of Computing, Macquarie University, Australia}~~~~\textit{{$^{*}$} Corresponding author}} \\
\{zhufeng.zhu, mingjie.zmj, xinxing.yangxx, longyao.llf, bruceyu.yl, zhangtiehua.zth, jun.zhoujun\}@antgroup.com, \\\{zjuccc, wufei\}@zju.edu.cn, \{guanfeng.liu, yan.wang\}@mq.edu.au
}

\maketitle

\begin{abstract}
In recommendation scenarios, there are two long-standing challenges, i.e., selection bias and data sparsity, which lead to a significant drop in prediction accuracy for both Click-Through Rate (CTR) and post-click Conversion Rate (CVR) tasks. To cope with these issues, existing works emphasize on leveraging Multi-Task Learning (MTL) frameworks (Category 1) or causal debiasing frameworks (Category 2) to incorporate more auxiliary data in the entire exposure/inference space $\mathcal{D}$ or debias the selection bias in the click/training space $\mathcal{O}$. However, these two kinds of solutions cannot effectively address the not-missing-at-random problem and debias the selection bias in $\mathcal{O}$ to fit the inference in $\mathcal{D}$. To fill the research gaps, we propose a \textbf{D}irect entire-space \textbf{C}ausal \textbf{M}ulti-\textbf{T}ask framework, namely DCMT, for post-click conversion prediction in this paper. Specifically, inspired by users' decision process of conversion, we propose a new counterfactual mechanism to debias the selection bias in $\mathcal{D}$, which can predict the factual CVR and the counterfactual CVR under the soft constraint of a counterfactual prior knowledge. Extensive experiments demonstrate that our DCMT can improve the state-of-the-art methods by an average of $1.07\%$ in term of CVR AUC on the offline datasets and $0.75\%$ in term of PV-CVR on the online A/B test (the Alipay Search). Such improvements can increase millions of conversions per week in real industrial applications, e.g., the Alipay Search.
\end{abstract}


\input{sections/introduction}
\input{sections/preliminary}
\input{sections/ourmethod}
\input{sections/experiment}

\input{sections/relatedwork}
\input{sections/conclusion}

\bibliographystyle{IEEEtran}
\bibliography{ICDE2023}

\end{document}

%% file: sections/introduction.tex
\section{Introduction} \label{Section_Introduction}
Recommender systems (RSs) have been proven to have a powerful filtering capability for users to obtain matched products or services in many industrial applications, e.g., e-commerce platforms \cite{ma2018entire,wen2020entire}, search engines \cite{zhang2014sequential}, payment platforms \cite{richardson2007predicting,wang2022escm}, social networking \cite{zhou2019deep}, video-sharing systems \cite{li2021dual}, and advertising \cite{pan2018field,xi2021modeling}. In these recommender systems, Click-Through Rate (CTR) and post-click Conversion Rate (CVR) predictions are considered as the two most fundamental tasks, which can help online marketplaces understand the underlying logic of their users' click and purchase behaviour. In this paper, we mainly focus on the post-click CVR task.
\begin{figure}[t]
	\begin{center}
		\includegraphics[width=0.48\textwidth]{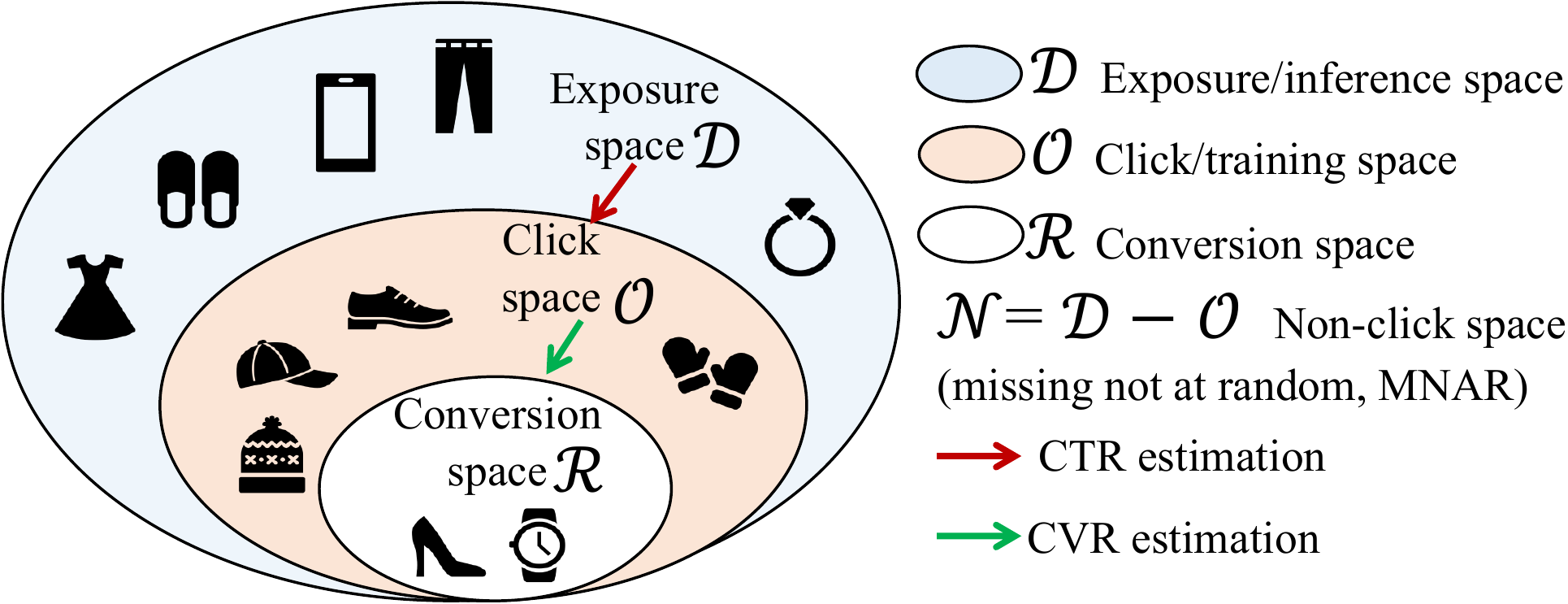}
		\vspace{-2mm}
		\caption{An example of the data sparsity and selection bias of the CVR estimation task in information-overloaded scenarios, where the training space $\mathcal{O}$ only contains clicked samples, while the inference space is the entire space $\mathcal{D}$ for all exposure samples.}
		\label{Challenges}
	\end{center}
	\vspace{-5mm}
\end{figure}

\subsection{Challenges}
In the above-mentioned recommendation scenarios, users tend to follow a general behaviour path, i.e., ``exposure$\rightarrow$click $\rightarrow$conversion", meaning both click and conversion labels are recorded in a training sample for RSs \cite{ma2018entire}. However, for the CVR task, there are two long-standing challenges, i.e., data sparsity and selection bias, which may result in the discrepancy between offline training performance and online testing metrics in real industrial applications.

\nosection{Data Sparsity}
Fig. \ref{Challenges} depicts a classical recommendation scenario which contains several exposed samples, a few clicked samples, and very few conversed samples. There are not enough clicked and/or conversed samples for recommender systems to train CTR and/or CVR prediction models. Such a phenomenon is not uncommon in e-commerce recommendation scenarios and referred to as data sparsity. For example, in the benchmark dataset  Ali-CCP\footnote{Dataset URL: https://tianchi.aliyun.com/dataset/dataDetail?dataId=408} from Alibaba\footnote{Alibaba website: https://www.alibaba.com}\cite{ma2018entire}, around 3.75\% of exposed items are clicked, and only 0.025\% of them are conversed (see the statistic in Table \ref{Datasets}). 
\begin{figure}[t]
 	\begin{center}
 	 	\subfigure[\textbf{Parallel MTL approaches} tend to parallelly design CTR estimation task and CVR estimation task and leverage the two auxiliary tasks, i.e., CTR task trained over $\mathcal{D}$ and CTCVR task trained over $\mathcal{D}$, to indirectly predict CVR.]{
 		\label{fig:subfig:solution1}
 		\includegraphics[width=0.48\textwidth]{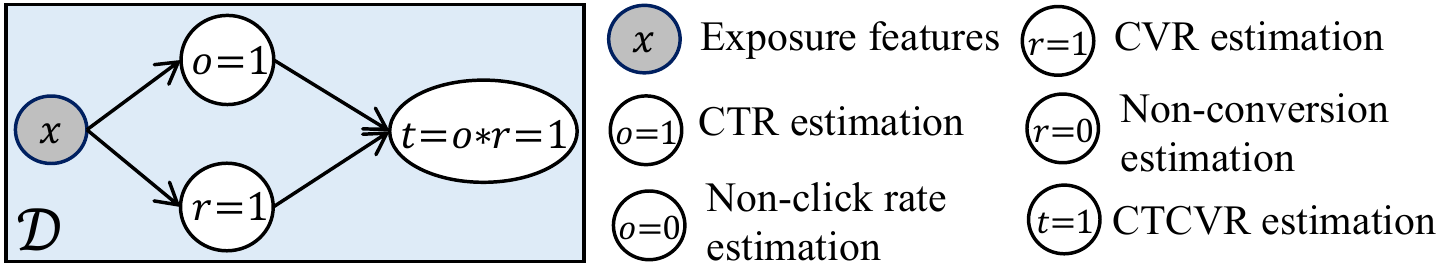}}
 		 \subfigure[\textbf{Multi-gate MTL approaches} tend to transfer the knowledge across the CTR task trained over $\mathcal{D}$ and the CVR task trained over $\mathcal{O}$.]{
 		\label{fig:subfig:solution2}
 		\includegraphics[width=0.23\textwidth]{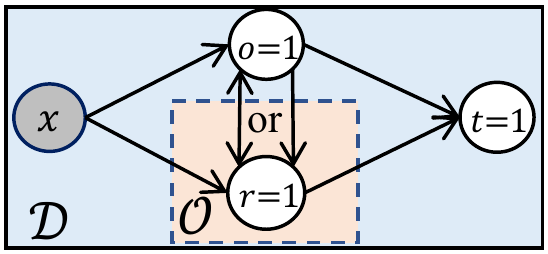}}
 		\hfill
 		\subfigure[\textbf{Propensity-based causal approaches} tend to eliminate the causal effect of input features $x$ (i.e., click propensity $p$) on the CTR task and debias the selection bias in $\mathcal{O}$.]{
 		\label{fig:subfig:solution3}
 		\includegraphics[width=0.23\textwidth]{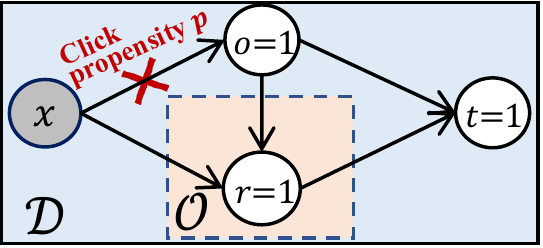}}
 		 \subfigure[Our \textbf{DCMT} aims to eliminate both the causal effect of $x$ (click propensity $p$) on the click event ($o=1$) and the causal effect of $x$ (non-click propensity $1-p$) on the non-click event ($o=0$), which can debias the selection bias in both the factual click space $\mathcal{O}$ and the counterfactual non-click space $\mathcal{N}^*$ (based on a counterfactual mechanism) for the CVR task.]{
 		\label{fig:subfig:our_idea}
 		\includegraphics[width=0.48\textwidth]{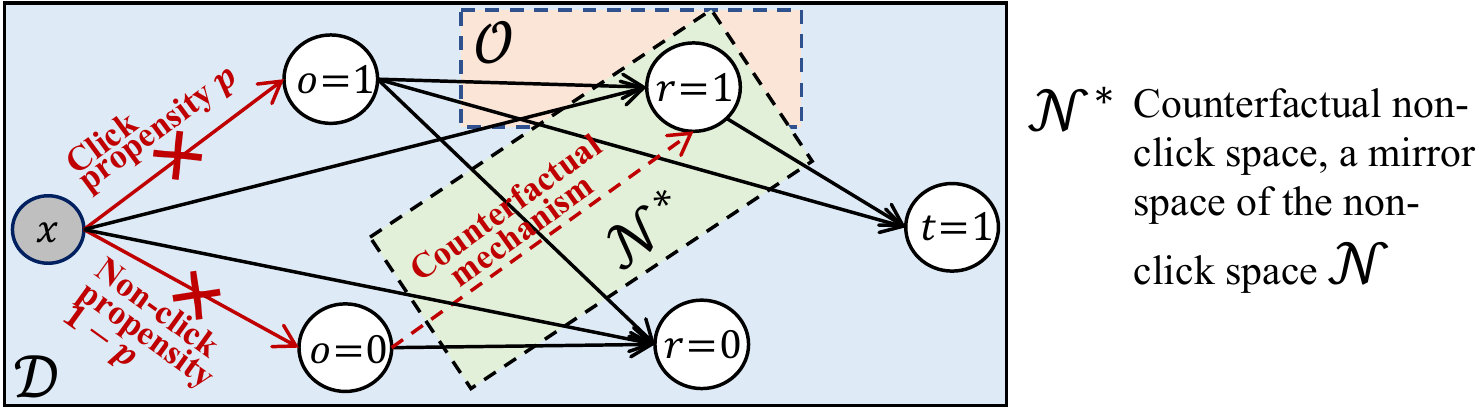}}
 		\vspace{-2mm}
 		\caption{Existing solutions and our idea.}
 	\label{the_existing_solutions_and_our_idea}
 	\end{center}
 	\vspace{-5mm}
 \end{figure}

\nosection{Selection Bias}\label{section_selectionbias}
Selection bias is a long-standing issue for conventional post-click CVR models \cite{ma2018entire,zhang2020large,wu2022opportunity}. As shown in Fig. \ref{Challenges}, for the conventional post-click CVR models, the data distribution in the training space $\mathcal{O}$ (the click space) is drawn but different from that in target population $\mathcal{D}$ (inference/exposure space). This is because the training data is biased by the user's self-selection. From the perspective of statistics, the missing data, e.g., exposed\&unclicked samples in the non-click space $\mathcal{N}$, is not missing at random (NMAR) \cite{zhang2020large,wu2022opportunity}. Users tend to click the items they like and thus items with lower CVR are less likely to be observed in $\mathcal{O}$. 

\subsection{Existing Solutions and Their Limitations}
\nosection{Parallel MTL approaches} To tackle the above challenges, i.e., data sparsity and selection bias, some of the existing studies (see Fig. \ref{fig:subfig:solution1}) tend to treat CTR prediction and CVR prediction as two parallel and related tasks in the same RS and share the input features by using Multi-Task Learning (MTL) frameworks, e.g., ESMM \cite{ma2018entire} and ESM$^2$ \cite{wen2020entire}. As shown in Fig. \ref{fig:subfig:solution1}, for the prediction task of post-click CVR, the MTL approaches \cite{ma2018entire, wen2020entire} tend to share the common embeddings of input features and train CTR and CVR models parallelly. The two auxiliary tasks, i.e., CTR prediction and the click\&conversion rate (CTCVR) prediction, are used to indirectly predict CVR via the probabilistic function $p(t=1|x)=p(o=1,r=1|x) = p(o=1|x)*p(r=1|x,o=1)$ \cite{ma2018entire}. The CTR and CTCVR tasks can be both trained on $\mathcal{D}$, which makes the training space the same as the inference space. This strategy can ease the problems of data sparsity and selection bias to some extent. However, as shown in Fig. \ref{fig:subfig:solution1}, these existing frameworks actually model another probabilistic function, i.e.,  $p(t=1|x) = p(o=1|x)*p(r=1|x)$. This means that these approaches ignore that the click event ($o=1$) has a causal effect on the conversion event ($r=1$) (\textbf{Limitation 1}), and thus fail to take advantage of such underlying correlations. This problem is also called potential independence priority (\textbf{PIP}) in the literature \cite{wang2022escm}.

\nosection{Multi-gate MTL approaches} To address \textbf{Limitation 1}, the other MTL models, e.g., Cross Stitch \cite{misra2016cross}, MMOE \cite{ma2018modeling}, PLE \cite{tang2020progressive}, MOSE \cite{qin2020multitask}, and AITM \cite{xi2021modeling}, are applied to build the relationship between the CTR task trained over $\mathcal{D}$ and the CVR task trained over $\mathcal{O}$. These multi-gate MTL approaches either apply some shared and/or specific expert modules to share the knowledge across the CTR and CVR tasks \cite{misra2016cross,ma2018modeling,tang2020progressive,qin2020multitask} or apply a mechanism to transfer the knowledge from the CTR task to the CVR task \cite{xi2021modeling} (see Fig. \ref{fig:subfig:solution2}). However, none of these MTL approaches attempt to address the problem of NMAR (\textbf{Limitation 2}).

\nosection{Causal approaches} To address \textbf{Limitation 2}, some causal approaches with multi-task learning, e.g., Multi-IPW/DR \cite{zhang2020large} and ESCM$^2$ \cite{wang2022escm}, are proposed to adapt for the data generation process and thus restore the information from NMAR data \cite{zhang2020large}. As shown in Fig. \ref{fig:subfig:solution3}, the core idea of these causal approaches is to eliminate the effect of input features $x$ (i.e., click propensity $p$) on the CTR task and thus obtain an conditionally unbiased estimation of CVR $p(r=1|do(o=1),x)$. The ``do" denotes the do-calculus that are applied to address the confounding bias in causal inference, i.e., the input features $x$ affect both the treatment $o=1$ and the outcome $r=1$ \cite{wu2022opportunity}. The existing causal approaches mainly focus on debiasing the selection bias in $\mathcal{O}$, e.g., Multi-IPW and ESCM$^2$-IPW, or leveraging a new auxiliary imputation task trained over $\mathcal{D}$ to improve the debiasing performance in $\mathcal{O}$, e.g., Multi-DR and ESCM$^2$-DR. The authors in Multi-DR and ESCM$^2$-DR believe that their strategy can indirectly debias the selection bias in $\mathcal{D}$. However, it is still hard to guarantee that the unbiased CVR estimation trained over $\mathcal{O}$ can infer very well over $\mathcal{D}$ (\textbf{Limitation 3}, see the Result 3-2 in Section \ref{online_experimental_results}).

\subsection{Our Approach and Contribution}
To further address \textbf{Limitation 3}, it is necessary to directly debias the CVR task of the causal models over $\mathcal{D}$ rather than only debias over $\mathcal{O}$, e.g., Multi-IPW and ESCM$^2$-IPW, or indirectly debias over $\mathcal{D}$, e.g., Multi-DR and ESCM$^2$-DR. However, in real application scenarios, there are not positive samples for the CVR task, i.e., the ``non-click$\rightarrow$conversion" samples, in the non-click space $\mathcal{N}$. These ``non-click$\rightarrow$conversion" samples are actually hidden in the true negative (``non-click$\rightarrow$non-conversion") samples as fake negative samples. $\mathcal{D}$ contains a lot of fake negative samples because users' attitude toward their unclicked items is not entirely negative. It could be that the users have not been aware of these unclicked items because of exposure position, display style, and other factors. Suppose the unclicked items are clicked, it is still possible for the users to purchase the non-click items. This means that blindly utilizing all unclicked samples as negative samples in $\mathcal{N}$ may lead to a discrepancy between users' conversion preference and their CVR predictions. Instead, in this paper, we propose a \textbf{D}irect entire-space \textbf{C}ausal \textbf{M}ulti-\textbf{T}ask framework, namely DCMT, for post-click conversion estimation. Our DCMT designs a new counterfactual mechanism (the underlying rationale will be analysed in Section \ref{section_counterfactual_mechanism}) to debias the selection bias in the counterfactual non-click space $\mathcal{N}^*$ (a mirror space of $\mathcal{N}$, see Fig. \ref{fig:subfig:our_idea}). The main contributions of our DCMT framework are summarized as follows.
\begin{itemize}[leftmargin=*]
    \item This is the first causal work that attempts to estimate the CVR by directly debiasing the selection bias in $\mathcal{D}$ ($\mathcal{O}$ + $\mathcal{N}^*$) rather than only debiasing the selection bias in $\mathcal{O}$. A theoretical proof of the unbiased CVR estimation of our DCMT is provided in this paper.
    \item Inspired by the decision process of conversion, we propose a new counterfactual mechanism to predict the factual CVR in the factual space and the counterfactual CVR in the counterfactual space under the soft constraint of a counterfactual prior knowledge. With the help of our proposed twin tower, the CVR estimation of our DCMT can effectively simulate users' decision process of conversion.
    \item We conduct extensive offline experiments on five real-world public datasets and an online A/B test on the Alipay Search system to verify the claims mentioned above. The experiments demonstrate that our DCMT can improve the state-of-the-art methods by an average of $1.07\%$ in term of CVR AUC on the offline datasets and $0.75\%$ in term of PV-CVR on the online A/B test. Such improvements can increase millions of conversions per week in real industrial applications, e.g., the Alipay Search.
\end{itemize}

%% file: sections/preliminary.tex
\section{Preliminaries}
In this section, we first formalize the definitions of CTR, CVR, and CTCVR. Then, to design an MTL framework with a reasonable knowledge transfer potential among these three tasks, we analyse the state-of-the-art MTL frameworks in practice, e.g., ESMM \cite{ma2018entire}. Finally, to obtain an unbiased CVR estimation over the entire exposure space $\mathcal{D}$, we introduce the core ideas of the state-of-the-art causal MTL approaches, i.e., MTL-IPW estimator and MTL-DR estimator \cite{zhang2020large, wang2022escm}. For the sake of better readability, we list the important notations of this paper in Table \ref{Notations}.

\begin{table}
		\caption{Important notations} \label{Notations}
		\vspace{-6mm}
		\begin{center}
				\resizebox{0.49\textwidth}{!}{
			\begin{tabular}{|c|c|}
				\hline
				\textbf{Symbol} & \textbf{Definition}\\
				\hline
				$\mathcal{U}=\{u_1,...,u_m\}$   & the set of $m$ users \\
				\hline
				$\mathcal{V}=\{v_1,...,v_n\}$ & the set of $n$ items\\
				\hline
				$x_{i,j}$ & \makecell{the input features of a sample (the user-item\\ pair $<u_i, v_j>$), including user features,\\ item features, and other context features}\\
				\hline
				$\mathcal{D}=\mathcal{U}\times\mathcal{V}$ & \makecell{the exposure space, the set of all exposed\\ user-item pairs, $d_{i,j} \in \{0,1\}$ indicates\\ whether the item $v_j$ is exposed to the user $u_i$}\\
				\hline
				\makecell{$\mathcal{O} = \{(i,j)|o_{i,j}=1,$\\$(i,j)\in \mathcal{D}\}$; $\mathcal{O} \in \mathbb{R}^{m \times n}$}& \makecell{the click space, $O$ is the click matrix of users\\ on items, $o_{i,j} \in \{0,1\}$ indicates whether\\ the user $u_i$ click on the item $v_j$}\\
				\hline
				\makecell{$\mathcal{N} = \{(i,j)|o_{i,j}=0,$\\$(i,j)\in \mathcal{D}\}$; $\mathcal{O} \in \mathbb{R}^{m \times n}$}& \makecell{the non-click space}\\
				\hline
				\makecell{$\mathcal{R} = \{(i,j)|r_{i,j}=1,$\\$(i,j)\in \mathcal{D}\}$; $\mathcal{R} \in \mathbb{R}^{m \times n}$} & \makecell{the conversion space, $R$ is the conversion\\ matrix of users on items, $r_{i,j} \in \{0,1\}$ indicates\\ whether the user $u_i$ purchases the item $v_j$}\\
				\hline
				$\hat{*}$ & \makecell{the predicted notations, e.g., $\hat{r}_{i,j}$ represents\\ the predicted CVR of $u_i$ on item $v_j$ }\\
				\hline
			\end{tabular}
		}
		\end{center}
		\vspace{-4mm}
\end{table}
%

\subsection{Unbiased Estimation}
First, if we can obtain the fully observed conversion labels, i.e., the conversion matrix $R$ is fully observed, the ideal loss function of the CVR prediction task (ground truth) can be formulated as:
\begin{equation}\label{Equation_cvr_loss_groundtruth}
\small{
	\begin{aligned}
	&\mathcal{E}^{\textrm{ground-truth}} = \mathcal{E}(R, \hat{R}) = \frac{1}{|\mathcal{D}|} \sum \limits_{(i,j) \in \mathcal{D}} e(r_{i,j},\hat{r}_{i,j}),
	\end{aligned}
	}
\end{equation}
where $e(r_{i,j},\hat{r}_{i,j})$ is the log loss, i.e., $e(r_{i,j},\hat{r}_{i,j})=-r_{i,j}log(\hat{r}_{i,j})-(1-r_{i,j})log(1-\hat{r}_{i,j})$.

However, in real-world applications, we can only observe a part of conversion labels, i.e., $R^{\textrm{obs}}$, in the click space $\mathcal{O}$, and the others, i.e.,  $R^{\textrm{mis}}$, are missing in the non-click space $\mathcal{N}$. Therefore, naive CVR estimators tend to be trained over $\mathcal{O}$ and their loss function can be formulated as:
\begin{equation}\label{Equation_cvr_loss_traditional}
\small{
	\begin{aligned}
	\mathcal{E}^{\textrm{naive}} &= \mathcal{E}(R^{\textrm{obs}}, \hat{R}) \\
	&= \frac{1}{|\mathcal{O}|} \sum \limits_{(i,j) \in \mathcal{O}} e(r_{i,j},\hat{r}_{i,j})\\
	&= \frac{1}{|\mathcal{O}|} \sum \limits_{(i,j) \in \mathcal{D}} o_{i,j}e(r_{i,j},\hat{r}_{i,j}).
	\end{aligned}
	}
\end{equation}

The bias of CVR loss between the naive CVR estimators and the ground truth is:
\begin{equation}\label{Equation_cvr_loss_ground_truth}
	\small{
	\begin{aligned}
	&\textrm{Bias}^{\textrm{naive}}=|\mathcal{E}^{\textrm{naive}} - \mathcal{E}^{\textrm{ground-truth}}| \\
	&=\bigg|\frac{1}{|\mathcal{O}|} \sum \limits_{(i,j) \in \mathcal{D}} o_{i,j}e(r_{i,j},\hat{r}_{i,j})-\frac{1}{|\mathcal{D}|} \sum \limits_{(i,j) \in \mathcal{D}} e(r_{i,j},\hat{r}_{i,j})\bigg|
	\end{aligned}
	}
\end{equation}

Finally, based on the above-mentioned bias of CVR loss, we formalize the definition of unbiased estimation as follows.
\begin{myDef}
\textbf{Unbiased Estimation of the CVR Prediction}: The CVR estimation of a model $M$ is unbiased when the expectation of the estimated loss in $\mathcal{O}$ equals the loss of the ground truth, i.e., $Bias^{M} = |E_{\mathcal{O}}(\mathcal{E}^{M}) - \mathcal{E}^{\textrm{ground-truth}}| = 0$.
\end{myDef}
As introduced in Section \ref{Section_Introduction}, if data is MNAR, then the data distribution of $\mathcal{O}$ is different from that of $\mathcal{D}$, and thus $\textrm{Bias}^{M} \gg 0$ (see Eq.(\ref{Equation_cvr_loss_traditional})), i.e., the CVR estimation of $M$ is biased.

\subsection{Multi-Task Learning for CVR Estimation}
CTR prediction and CTCVR prediction are auxiliary tasks for the main task, i.e., CVR estimation. The prediction results of CTR and CTCVR can be used to indirectly obtain the prediction result of CVR. This strategy of MTL is widely used in the literature, such as ESMM \cite{ma2018entire} and ESM$^2$ \cite{wen2020entire}. The underlying reason behind this strategy is that the traditional CVR prediction task can be only trained on $\mathcal{O}$, i.e., all clicked samples, shown in Fig. \ref{Challenges}, and the clicked samples are very sparse. Thus, with the help of the CTR prediction task and CTCVR prediction task (trained on $\mathcal{D}$ with all exposed items shown in Fig. \ref{Challenges}), the problems of data sparsity and selection bias are expected to be alleviated to some extent.  

However, these parallel MTL approaches cannot obtain an unbiased CVR estimation. We choose the representative method, i.e., ESMM, as an example to demonstrate that the CVR estimation of these parallel MTL approaches is biased, which has been theoretically analysed in the literature \cite{zhang2020large}. 

In addition to the parallel MTL methods, e.g., ESMM, many multi-gate MTL methods, e.g., Cross Stitch \cite{misra2016cross}, MMOE \cite{ma2018modeling}, PLE \cite{tang2020progressive}, MOSE \cite{qin2020multitask}, and AITM \cite{xi2021modeling}, are proposed to build the relationship between the CTR task and the CVR task. However, none of them attempts to address the problem of NMAR, and thus they cannot guarantee that the CVR estimations derived from their methods are unbiased.

\subsection{Propensity-based Debiasing for CVR Estimation} \label{section_ipw}
As introduced in Section \ref{Section_Introduction}, the click propensities of users on items, i.e., $P$, leads to the distribution difference between $\mathcal{O}$ and $\mathcal{D}$. Thus, some inverse propensity weighting (IPW)-based approaches, e.g., Multi-IPW \cite{zhang2020large} and ESCM$^2$-IPW \cite{wang2022escm}, are proposed to eliminate the effect of the input features $x$ on users' click events. The loss function of IPW-based CVR estimators can be formulated as:
\begin{equation}\label{Equation_cvr_loss_ipw}
\small{
	\begin{aligned}
	\mathcal{E}^{\textrm{IPW}} =\frac{1}{|\mathcal{D}|} \sum \limits_{(i,j) \in \mathcal{D}} \frac{o_{i,j}e(r_{i,j},\hat{r}_{i,j})}{\hat{p}_{i,j}},
	\end{aligned}
	}
\end{equation}
where $\hat{p}_{i,j}$ is the prediction of click propensity of user $u_i$ on item $v_j$. Intuitively, it can eliminate the effect of click propensity on the CVR estimation by giving an inverse propensity weight, i.e., $1/\hat{p}_{i,j}$, to the corresponding CVR loss, i.e., $e(r_{i,j},\hat{r}_{i,j})$. Note that the click propensity of user $u_i$ on item $v_j$ is the CTR of $u_i$ on item $v_j$, i.e., $\hat{p}_{i,j}=\hat{o}_{i,j}$. Therefore, some existing causal approaches, e.g., Multi-IPW and ESCM$^2$-IPW, leverage multi-task learning to simultaneously learn an auxiliary propensity prediction, i.e., CTR prediction, with CVR prediction. These approaches are categorised into the group of MTL-IPW. The corresponding loss function of MTL-IPW-based CVR estimators can be rewritten as follows. 
\begin{equation}\label{Equation_cvr_loss_ipw_2}
\small{
	\begin{aligned}
	\mathcal{E}^{\textrm{MTL-IPW}} &=\frac{1}{|\mathcal{D}|} \sum \limits_{(i,j) \in \mathcal{D}} \frac{o_{i,j}e(r_{i,j},\hat{r}_{i,j})}{\hat{o}_{i,j}}\\
	&=\frac{1}{|\mathcal{D}|} \sum \limits_{(i,j) \in \mathcal{O}} \frac{e(r_{i,j},\hat{r}_{i,j})}{\hat{o}_{i,j}}.
	\end{aligned}
	}
\end{equation}

\nosection{MTL-IPW Analysis} The precondition of unbiased estimation of IPW-based approaches is that the prediction of click propensity (CTR prediction) is accurate. However, in practice, it is almost impossible for these IPW-based approaches to guarantee this precondition in the training process. Also, as shown in Eq.(\ref{Equation_cvr_loss_ipw_2}), the IPW-based CVR approaches only consider the CVR loss in $\mathcal{O}$, i.e., $o_{i,j}=1$. This means that these IPW-based CVR approaches are actually trained over $\mathcal{O}$ rather than $\mathcal{D}$.
\subsection{Doubly Robust for CVR Estimation}\label{section_dr}
 To address the above-mentioned issues of MTL-IPW approaches, some doubly robust-based approaches, e.g., Multi-DR \cite{zhang2020large} and ESCM$^2$-DR \cite{wang2022escm}, are proposed in the literature \cite{zhang2020large,wang2022escm}. In addition to CTR task and CVR task, these doubly robust-based approaches also propose an imputation task trained over $\mathcal{D}$ to estimate the CVR error $e_{i,j}=e(r_{i,j},\hat{r}_{i,j})$ with $\hat{e}_{i,j}$. The performance of this imputation task is assessed by $\delta_{i,j}=e_{i,j}-\hat{e}_{i,j}$. Similarly, these approaches are categorised in the group of MTL-DR. The loss function of MTL-DR-based CVR estimators can be formulated as follows. 
\begin{equation}\label{Equation_cvr_loss_dr}
\small{
	\begin{aligned}
	\mathcal{E}^{\textrm{MTL-DR}} =\frac{1}{|\mathcal{D}|} \sum \limits_{(i,j) \in \mathcal{D}} \big( \hat{e}_{i,j} + \frac{o_{i,j}\delta_{i,j}}{\hat{o}_{i,j}}\big).
	\end{aligned}
	}
\end{equation}

\nosection{MTL-DR Analysis}\label{section_DR_analysis} As indicated in Eq. (\ref{Equation_cvr_loss_dr}), the first term, i.e., $\hat{e}_{i,j}$ (the prediction of CVR error), occurs in $\mathcal{D}$. However, in $\mathcal{N} \in \mathcal{D}$, the conversion label $r_{i,j}$ is not accurate, i.e., there are fake negative samples in this space. The second term, i.e., $\frac{o_{i,j}\delta_{i,j}}{\hat{o}_{i,j}}$ (the weighted loss of the imputation task), only occurs in $\mathcal{O}$. Also, compared with MTL-IPW, the precondition of unbiased estimation of DR-based approaches becomes two `or' conditions, i.e., the CTR predictions are accurate or the predictions of CVR loss are accurate. However, it is difficult to convince us that the main task, i.e., CVR task, with the two uncertain auxiliary tasks, i.e., CTR task and imputation task, is better than that with the one uncertain auxiliary task, i.e., CTR task.

%% file: sections/ourmethod.tex
\section{The Proposed Model}\label{Section_OurMethod}
 \begin{figure*}[t]
	\begin{center}
		\includegraphics[width=0.75\textwidth]{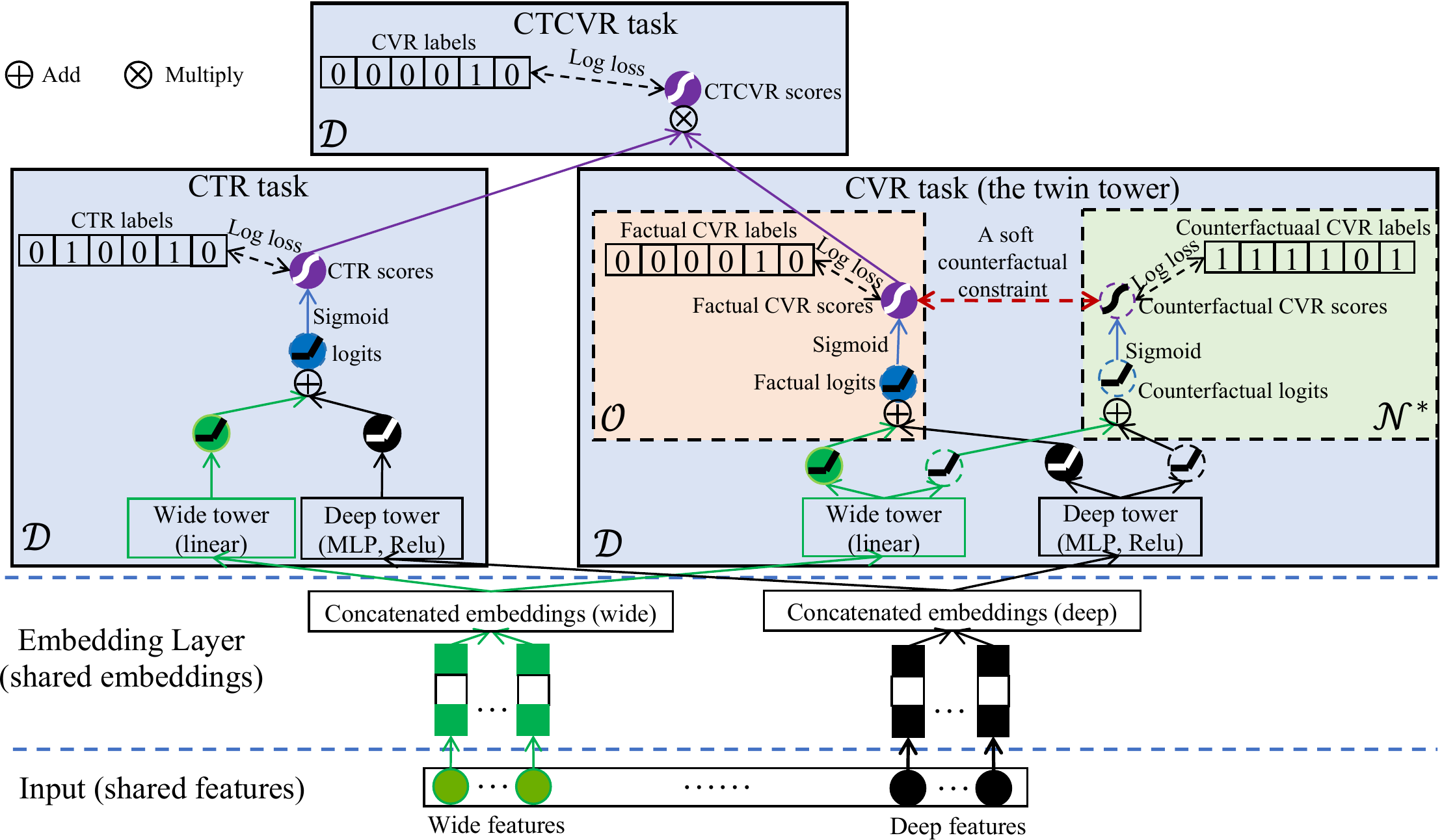}
		\vspace{-2mm}
		\caption{The structure of our DCMT model.}
		\label{our_model_structure}
	\end{center}
	\vspace{-5mm}
\end{figure*}
All the above-mentioned issues of the existing CVR approaches motivate us to propose our \textbf{D}irect entire-space \textbf{C}ausal \textbf{M}ulti-\textbf{T}ask framework, namely DCMT, for post-click conversion estimation scenarios.

\subsection{Overview of Our Framework}\label{Section_overview}
We briefly present each component of DCMT as follows.

\nosection{Input (shared features)} First, for the input of our DCMT, we classify the input features into two categories: deep features and wide features. In \emph{CTR Task} and \emph{CVR Task}, we can combine the benefits of generalization (trained on deep features) and memorization (trained on wide features)  \cite{cheng2016wide}. In this paper, we collect the features of user profiles and item details, e.g., gender (user), age (user), and shop id (item), as the deep features, which can be trained to learn the deep and complex relationship between users and items. We collect the interaction features, e.g., favourite shop id (from user to item), which can be trained to understand users' historical preferences. Note that if a training dataset does not contain any wide features, our DCMT framework will degenerate from a wide\&deep structure to a pure deep structure.

\nosection{Embedding Layer} In this layer, the input features are embedded into low-dimension spaces and we concatenate the embeddings of deep features,  and wide features, respectively. The concatenated deep embeddings and wide embeddings are shared by \emph{CTR Task} and \emph{CVR Task}. 

\nosection{CTR Task} In this component, we adopt a Multi-Layer Perceptron (MLP), i.e., CTR tower (deep part) in Fig. \ref{our_model_structure}, to represent the deep relation between users and items. Also, we adopt a generalized linear structure, i.e., CTR tower (wide part), which can be trained to capture users' historical preferences over $\mathcal{D}$. Based on these two CTR towers, we can obtain their corresponding outputs, i.e., CTR predictions.

\nosection{CVR Task} Like \emph{CTR Task}, the wide and deep concatenated embeddings are fed into the twin tower of \emph{CTR Task}. In this task, we will propose a new counterfactual mechanism to predict the factual CVR and the counterfactual CVR simultaneously. These two CVR predictions are generated by the twin tower (see its structure in Fig. \ref{our_model_structure}) that will be introduced in Section \ref{section_twin_tower}. The two predictions follow a counterfactual prior knowledge, i,e, the sum of them should be $1$. With the help of the counterfactual mechanism, our CVR estimator can leverage all the samples in $\mathcal{D}$, which can address the problem of data sparsity. Compared with the traditional propensity-based debiasing approaches, our counterfactual mechanism can also help our DCMT debias the selection bias in the counterfactual non-click space $\mathcal{N}^*$. Note that the counterfactual CVR task is an auxiliary task for the factual CVR task, and we only retain the predictions of the factual CVR task as the final results of the CVR task.

\nosection{CTCVR Task} Finally, we can obtain the CTCVR predictions over $\mathcal{D}$ by multiplying the CTR predictions and the CVR predictions, i.e., $\hat{t}_{i,j} = \hat{o}_{i,j}*\hat{r}_{i,j}$.

 Since the CVR task is our main task, and thus, we will introduce the details of \emph{CVR Task} in the following sections.

\subsection{CVR Estimation over the Entire Exposure Space $\mathcal{D}$}
As introduced in Section \ref{section_ipw}, the IPW-based approaches focus on debiasing the selection bias caused by click propensity in $\mathcal{O}$. However, these approaches seem to ignore the samples in $\mathcal{N}$ in the training of CVR models. To leverage all the samples in $\mathcal{D}$ and alleviate the problems of selection bias and data sparsity in $\mathcal{D}$, the CVR loss function of naive solutions is formulated as follows.
\begin{equation}\label{Equation_cvr_loss_our_method_naive}
\small{
	\begin{aligned}
	\mathcal{E}^{\textrm{DCMT\_naive}} &=\frac{1}{|\mathcal{D}|} \bigg( \sum \limits_{(i,j) \in \mathcal{O}} \frac{e_{i,j}}{\hat{o}_{i,j}} + \sum \limits_{(i,j) \in \mathcal{N}} \frac{e_{i,j}}{1-\hat{o}_{i,j}} \bigg)\\
	&=\frac{1}{|\mathcal{D}|} \sum \limits_{(i,j) \in \mathcal{D}} \left( \frac{o_{i,j}e_{i,j}}{\hat{o}_{i,j}} + \frac{(1-o_{i,j})e_{i,j}}{1-\hat{o}_{i,j}} \right),
	\end{aligned}
	}
\end{equation}
where $e_{i,j}=e(r_{i,j},\hat{r}_{i,j})$. In contrast to the click propensity, i.e., $\hat{p}_{i,j}$, in $\mathcal{O}$, the non-click propensity, i.e., $1-\hat{p}_{i,j}$, in $\mathcal{N}$ also leads to a selection bias. However, in $\mathcal{N}$, the conversion labels are always $0$, i.e., $r_{i,j}=0, \forall(i,j) \in \mathcal{N}$. There are fake negative samples in $\mathcal{N}$. This is the main reason that the conventional CVR estimators did not leverage the samples in $\mathcal{N}$ to reduce the selection bias. Therefore, we cannot directly apply Eq. (\ref{Equation_cvr_loss_our_method_naive}) to reduce the selection bias caused by click propensity in $\mathcal{D}$.

\subsection{Counterfactual Mechanism}\label{section_counterfactual_mechanism}
As introduced in Section \ref{section_dr}, the DR-based approaches apply an auxiliary imputation task to predict the CVR error, i.e., $\hat{e}_{i,j}$. This imputation tower is trained over $\mathcal{D}$, and thus it is expected to reduce the selection bias in $\mathcal{D}$. However, we have analysed its disadvantages in Section \ref{section_DR_analysis}. To address these disadvantages, in this section, we propose another promising solution, i.e., a new counterfactual mechanism, to achieve the same goal of training over $\mathcal{D}$.
\begin{figure}[t]
	\begin{center}
		\includegraphics[width=0.4\textwidth]{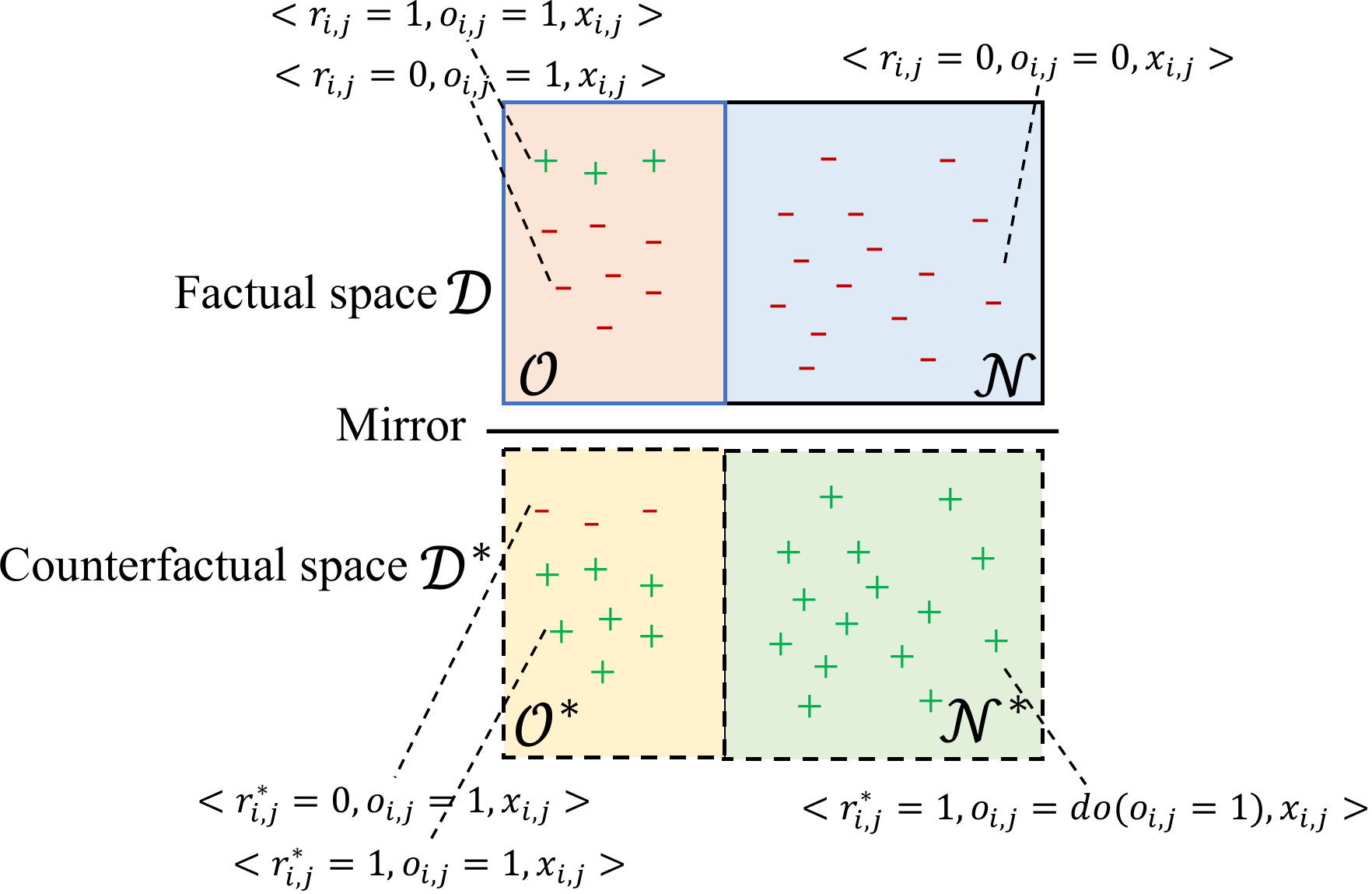}
		\vspace{-2mm}
		\caption{The counterfactual space is the mirror space of the corresponding factual space. A counterfactual sample is its corresponding factual sample with an opposite status (label).}
		\label{Counterfactual_space}
	\end{center}
	\vspace{-2mm}
\end{figure}
\begin{figure}[t]
	\begin{center}
		\includegraphics[width=0.35\textwidth]{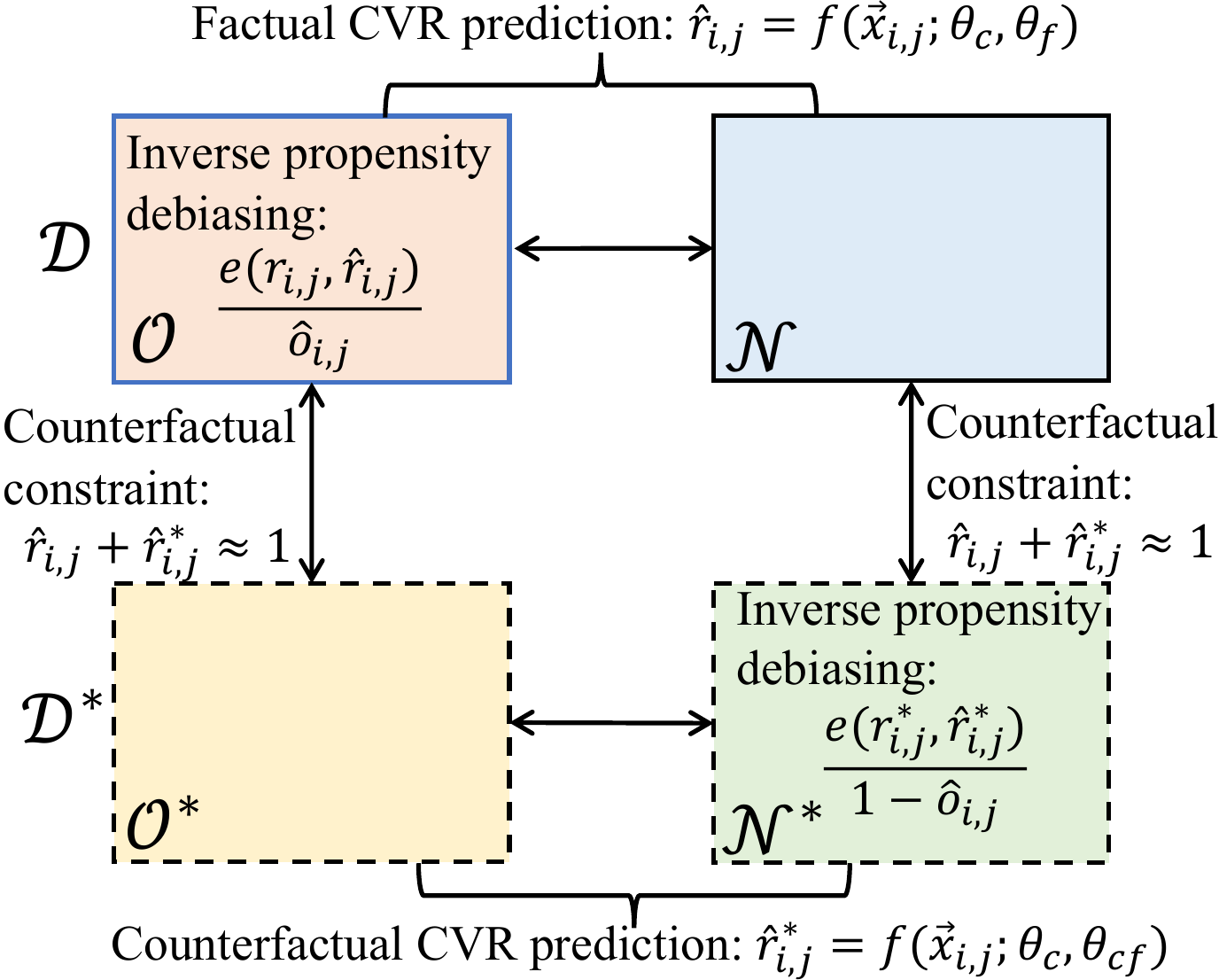}
		\vspace{-2mm}
		\caption{Debiasing circle. The debiasing strategies, i.e., $\frac{e(r_{i,j},\hat{r}_{i,j})}{\hat{o}_{i,j}}$ in $\mathcal{O}$ and  $\frac{e(r^*_{i,j},\hat{r}^*_{i,j})}{1-\hat{o}_{i,j}}$ in $\mathcal{N}^*$, can indirectly affect the CVR predictions in $\mathcal{O}^*$ and $\mathcal{N}$ respectively, via the soft counterfactual constraint, i.e., $\hat{r}_{i,j}+\hat{r}^*_{i,j} \approx 1$.}
		\label{debiasing_circle}
	\end{center}
	\vspace{-5mm}
\end{figure}

\nosection{Underlying Rationale}\label{underlying_rationale} Inspired by the decision process of conversion, we propose a concept of counterfactual sample space for CVR tasks (see Fig. \ref{Counterfactual_space}). From the perspective of probability, the user $u_i$ purchases the item $v_j$ with a probability of $p_{i,j}$ (conversion rate), and thus $u_i$ does not purchase $v_j$ with a probability of $1-p_{i,j}$. Suppose the decision of conversion is true, i.e., $r_{i,j}=1$, then there is a positive sample $<r_{i,j}=1, o_{i,j}=1, x_{i,j}>$ in the factual exposure $\mathcal{D}$. However, in this case, we cannot observe a negative sample $<r_{i,j}=0, o_{i,j}=1, x_{i,j}>$ though this counterfactual negative sample could have been observed with the probability of $1-p_{i,j}$. In our setting of counterfactual mechanism, this counterfactual negative sample will be included in the counterfactual exposure space $\mathcal{D}^*$. From the perspective of physics, in this paper, the counterfactual space $\mathcal{D}^*$ is the space mirror of the factual space $\mathcal{D}$. A counterfactual sample is its corresponding factual sample with an opposite status, which is similar to the relation between matter and antimatter \cite{canetti2012matter}.

\nosection{A Different Counterfactual Mechanism} The traditional counterfactual mechanism in causal inference tends to extend the factual/observed sample space. These counterfactual samples may have the same or similar features as the observed samples and their labels are predicted or generated by reasonable prior knowledge. However, this mechanism still cannot guarantee that the predicted/generated labels are accurate or the distribution of counterfactual samples is unbiased.

Since the definition of counterfactual samples in this paper is different from that in traditional causal inference \cite{morgan2015counterfactuals}, for readability, we formalize the definition of counterfactual samples as follows.
\begin{myDef}
\textbf{Counterfactual Samples}: A counterfactual sample is the mirror image of the corresponding factual sample, e.g., $<r^*_{i,j}=1-r_{i,j}, o_{i,j}, x_{i,j}>$ vs. $<r_{i,j}, o_{i,j}, x_{i,j}>$. 
\end{myDef}

As shown in Fig. \ref{Counterfactual_space}, for a unclicked\&unconversed sample $<r_{i,j}=0, o_{i,j}=0, x_{i,j}>$ in the non-click space $\mathcal{N}$, we can generate its corresponding counterfactual sample $<r^*_{i,j}=1, do(o_{i,j})=1, x_{i,j}>$ in $\mathcal{N}^*$. Here, the ``do" denotes the do-calculus that are applied to address the confounding bias in causal
inference. In this case, ``do" means that we suppose the unclicked\&unconversed sample is clicked, then we can predict its counterfactual CVR based on its corresponding counterfactual sample.

Therefore, under the setting of our counterfactual mechanism, the CVR loss function of our DCMT can be rewritten as follows.
\begin{equation}\label{Equation_cvr_loss_our_method}
\small{
	\begin{aligned}
	&\mathcal{E}^{\textrm{DCMT\_main}}\\
	&=\frac{1}{|\mathcal{D}|} \sum \limits_{(i,j) \in \mathcal{D}} \left( \frac{o_{i,j}e(r_{i,j},\hat{r}_{i,j})}{\hat{o}_{i,j}} + \frac{(1-o_{i,j})e(1-r_{i,j},\hat{r}^*_{i,j})}{1-\hat{o}_{i,j}} \right)\\
	&=\frac{1}{|\mathcal{D}|} \bigg( \sum \limits_{(i,j) \in \mathcal{O}} \frac{e(r_{i,j},\hat{r}_{i,j})}{\hat{o}_{i,j}} + \sum \limits_{(i,j) \in \mathcal{N}} \frac{e(1-r_{i,j},\hat{r}^*_{i,j})}{1-\hat{o}_{i,j}} \bigg)\\
	&=\frac{1}{|\mathcal{D}|} \bigg( \sum \limits_{(i,j) \in \mathcal{O}} \frac{e(r_{i,j},\hat{r}_{i,j})}{\hat{o}_{i,j}} + \sum \limits_{(i,j) \in \mathcal{N^*}} \frac{e(r^*_{i,j},\hat{r}^*_{i,j})}{1-\hat{o}_{i,j}} \bigg),
	\end{aligned}
	}
\end{equation}
where $\hat{r}^*_{i,j}$ is the corresponding prediction of counterfactual CVR (see Eq. (\ref{Equation_twin_tower_details})). Actually, in Eq. (\ref{Equation_cvr_loss_our_method}), the second term suffers from the problem of fake negative samples as well because we only reverse the conversion status of samples from $\mathcal{N}$ to $\mathcal{N}^*$. To address this issue, we employ a counterfactual regularizer in the following section.

\nosection{Counterfactual Prior Knowledge and Regularizer} 
As mentioned in \textbf{Underlying Rationale}, the user $u_i$ purchases the item $v_j$ with a probability of $p_{i,j}$ (conversion rate), and thus $u_i$ does not purchase $v_j$ with a probability of $1-p_{i,j}$. The prediction of factual CVR $\hat{r}_{i,j}$ and the prediction of counterfactual CVR $\hat{r}^*_{i,j}$ should also follow this prior knowledge, i.e., $\hat{r}_{i,j}+\hat{r}^*_{i,j}=1$. However, if we force $\hat{r}_{i,j}+\hat{r}^*_{i,j}$ to be $1$, then $\hat{r}_{i,j}$ and $\hat{r}_{i,j}$ will be restricted in respective small value ranges, e.g.,$[0.265, 0.305]$ and $[0.695, 0.735]$ (see Figure \ref{parameter_of_prediction_ranges} in Section \ref{section_impact_of_hp}). This kind of hard constraint makes the training process of our CVR estimation unable to minimize the factual loss and the counterfactual loss in Eq. (\ref{Equation_cvr_loss_our_method}). Therefore, the hard constraint is harmful to our CVR estimation. Instead, we add a counterfactual regularizer $L$ (a soft constraint) in Eq. (\ref{Equation_cvr_loss_our_method}) to minimize the error between $1$ and $\hat{r}_{i,j}+\hat{r}^*_{i,j}$ as follows.
\begin{equation}\label{Equation_cvr_loss_our_method_full}
\small{
	\begin{aligned}
	&\mathcal{E}^{\textrm{DCMT}} =\mathcal{E}^{\textrm{DCMT\_main}} + L\\
	 &=\frac{1}{|\mathcal{D}|} \Bigg( \underbrace{\sum \limits_{(i,j) \in \mathcal{O}} \frac{e(r_{i,j},\hat{r}_{i,j})}{\hat{o}_{i,j}}}_{\mathclap{\text{factual loss in } \mathcal{O}}} + \underbrace{\sum \limits_{(i,j) \in \mathcal{N^*}} \frac{e(r^*_{i,j},\hat{r}^*_{i,j})}{1-\hat{o}_{i,j}}}_{\mathclap{\text{counterfactual loss in }\mathcal{N^*}}} \Bigg)\\
	&+ \underbrace{\frac{\lambda_1}{|\mathcal{D}|}\sum \limits_{(i,j) \in \mathcal{D}} \left|1-(\hat{r}_{i,j}+\hat{r}^*_{i,j})\right|}_{\mathclap{\text{counterfactual regularizer}~L}},
	\end{aligned}
	}
\end{equation}
where $\lambda_1$ is the hyper-parameter to control the importance of the counterfactual regularizer. The counterfactual regularizer can help to eliminate the negative effect of the fake negative samples in $\mathcal{N}$, i.e., the fake positive samples in $\mathcal{N^*}$. As indicated in Eq. (\ref{Equation_cvr_loss_our_method_full}), the first term, i.e., factual loss in $\mathcal{O}$, is used to debias the selection bias in $\mathcal{O}$, while the second term, i.e., counterfactual loss in $\mathcal{N}^*$, is used to debias the selection bias in $\mathcal{N}^*$. 

With the help of the counterfactual regularizer, i.e., the third term in Eq. (\ref{Equation_cvr_loss_our_method_full}), the CVR task of our DCMT actually forms an interesting debiasing circle as shown in Fig. \ref{debiasing_circle}. The debiasing strategy, i.e., $\frac{e(r_{i,j},\hat{r}_{i,j})}{\hat{o}_{i,j}}$, in $\mathcal{O}$ directly affects the predictions of factual CVR in $\mathcal{N}$ via the shared prediction function, i.e., $\hat{r}_{i,j} = f(\vec{x}_{i,j}; \theta_{c}, \theta_{f})$, and indirectly affects the predictions of counterfactual CVR in $\mathcal{O}^*$ via the soft counterfactual constraint, i.e., $\hat{r}_{i,j}+\hat{r}^*_{i,j} \approx 1$. Similarly, the debiasing strategy, i.e., $\frac{e(r^*_{i,j},\hat{r}^*_{i,j})}{1-\hat{o}_{i,j}}$, in $\mathcal{N}^*$ directly affects the predictions of counterfactual CVR in $\mathcal{O}^*$ via the shared prediction function, i.e., $\hat{r}^*_{i,j} = f(\vec{x}_{i,j}; \theta_{c}, \theta_{cf})$, and indirectly affects the predictions of factual CVR in $\mathcal{N}$. This debiasing circle can achieve the goal of debiasing the selection bias in both $\mathcal{D}$ and $\mathcal{D}^*$, which can address the disadvantages of MTL-IPW and MTL-DR. The prediction functions of the factual CVR and the counterfactual CVR will be carefully introduced in the next section.
 \begin{figure}[t]
	\begin{center}
		\includegraphics[width=0.48\textwidth]{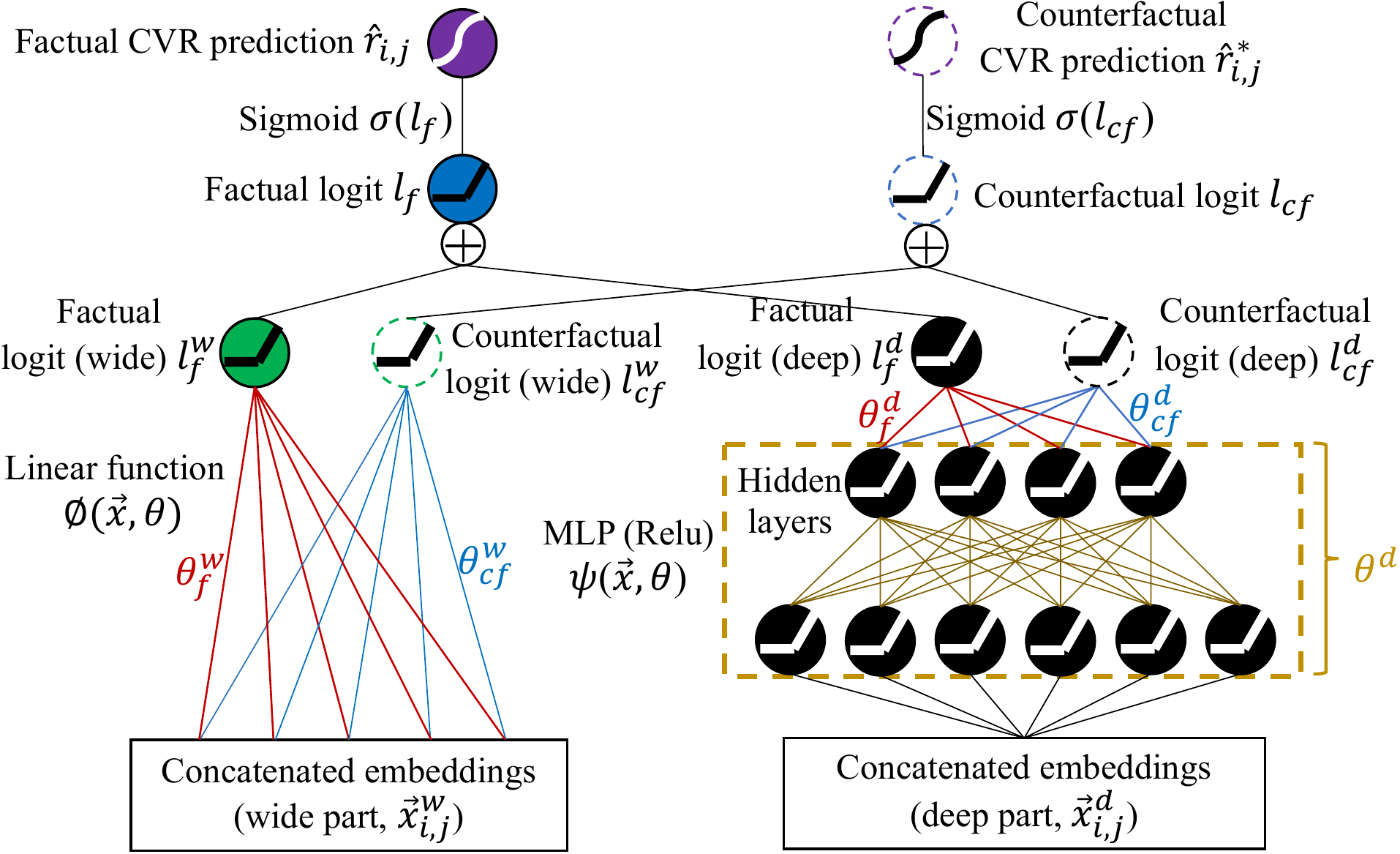}
		\vspace{-2mm}
		\caption{The detailed wide\&deep structure of our twin tower. In contrast to the general prediction function of our twin tower, i.e., Eq. (\ref{Equation_twin_tower}), $\theta_c = \theta^d$ (there are only the common parameters $\theta^d$ in the deep part), $\theta_f = \theta_f^w + \theta_f^d$, and $\theta_{cf} = \theta_{cf}^w + \theta_{cf}^d$. The detailed function is formulated in Eq. (\ref{Equation_twin_tower_details}).}
		\label{structure_twin_tower}
	\end{center}
	\vspace{-5mm}
\end{figure}
\subsection{Unbiasd CVR Estimation of Our DCMT}
\begin{myTheorem}
The CVR estimation of our DCMT is unbiased in the exposure space $\mathcal{D}$, i.e., Bias$^{\textrm{DCMT}}$ = $|E_{\mathcal{O}}(\mathcal{E}^{\textrm{DCMT}}) - \mathcal{E}^{\textrm{ground-truth}}| = 0$, when $o_{i,j}=\hat{o}_{i,j}$ and $\hat{r}_{i,j}+\hat{r}^*_{i,j}=1$, i.e., the prediction of click propensity is accurate and the predictions of factual CVR and counterfactual CVR follow the counterfactual prior knowledge.
\end{myTheorem}

Note that $o_{i,j}=\hat{o}_{i,j}$ means that $o_{i,j}=\hat{o}_{i,j}=1$ in the click space $\mathcal{O}$ and $o_{i,j}=\hat{o}_{i,j}=0$ in the non-click space $\mathcal{N}$. Meanwhile, $\hat{r}_{i,j}+\hat{r}^*_{i,j}=1$ means that the counterfactual regularizer $L=0$ and  $e(1-r_{i,j},\hat{r}^*_{i,j})=e(1-r_{i,j},1-\hat{r}_{i,j})=e(r_{i,j},\hat{r}_{i,j})$ because $e(r_{i,j},\hat{r}_{i,j})$ is the log loss.

\begin{proof}
\begin{equation}\label{DCMT_bias}
\small{
	\begin{aligned}
&\textrm{Bias}^{\textrm{DCMT}} = \left| \mathcal{E}^{\textrm{DCMT}} - \mathcal{E}^{\textrm{ground-truth}} \right|\\
 =&\bigg| \frac{1}{|\mathcal{D}|} \bigg( \sum \limits_{(i,j) \in \mathcal{O}} \frac{e(r_{i,j},\hat{r}_{i,j})}{\hat{o}_{i,j}} + \sum \limits_{(i,j) \in \mathcal{N^*}} \frac{e(r^*_{i,j},\hat{r}^*_{i,j})}{1-\hat{o}_{i,j}}\bigg)\\
	&+ L - \frac{1}{|\mathcal{D}|} \sum \limits_{(i,j) \in \mathcal{D}} e(r_{i,j},\hat{r}_{i,j})\bigg|\\
=&\bigg|\frac{1}{|\mathcal{D}|} \bigg( \sum \limits_{(i,j) \in \mathcal{O}} \frac{e(r_{i,j},\hat{r}_{i,j})}{\hat{o}_{i,j}} + \sum \limits_{(i,j) \in \mathcal{N}} \frac{e(1-r_{i,j},\hat{r}^*_{i,j})}{1-\hat{o}_{i,j}}\bigg)+ L\\
	& - \frac{1}{|\mathcal{D}|} \sum \limits_{(i,j) \in \mathcal{D}} e(r_{i,j},\hat{r}_{i,j})\bigg|\\
=&\bigg| \frac{1}{|\mathcal{D}|} \sum \limits_{(i,j) \in \mathcal{O}} \frac{(1-\hat{o}_{i,j})e(r_{i,j},\hat{r}_{i,j})}{\hat{o}_{i,j}} + L\\
&+ \frac{1}{|\mathcal{D}|}\sum \limits_{(i,j) \in \mathcal{N}} \frac{e(1-r_{i,j},\hat{r}^*_{i,j}) - (1-\hat{o}_{i,j})e(r_{i,j},\hat{r}_{i,j})}{1-\hat{o}_{i,j}}\bigg|=0.
	\end{aligned}
	}
\end{equation}
\end{proof}

\subsection{Twin Tower}\label{section_twin_tower}
As introduced in \textbf{Underlying Rationale} of Section \ref{section_counterfactual_mechanism}, our counterfactual mechanism is inspired by the decision process of conversion. To simulate users' decision process of conversion, we design a twin tower to predict the factual CVR and the counterfactual CVR simultaneously with the same input features. The general prediction function of the twin tower is formulated as follows.
\begin{equation}\label{Equation_twin_tower}
\small{
	\begin{aligned}
	&< \hat{r}_{i,j}, \hat{r}^*_{i,j} >= f(\vec{x}_{i,j}; \theta_{c}, \theta_{f}, \theta_{cf}),
	\end{aligned}
	}
\end{equation}
where $\vec{x}_{i,j}$ is the feature embedding vector of the input features $x_{i,j}$, $\theta_{c}$ are the common parameters for both $\hat{r}_{i,j}$ and $\hat{r}^*_{i,j}$, $\theta_{f}$ are the specific parameters for $\hat{r}_{i,j}$, and $\theta_{cf}$ are the specific parameters for $\hat{r}^*_{i,j}$. The common parameters represent the same thoughts for decision making when receiving the same input, while the specific parameters represent the divergent thoughts to make the final decisions, i.e., conversion or non-conversion. In the training process, the common parameters $\theta_{c}$ are updated by minimizing both the losses of the factual CVR task and the counterfactual CVR task. In contrast to the existing CVR estimators, e.g., MTL-IPW and MTL-DR, this strategy may avoid that the parameters of CVR estimators are biased to the samples in $\mathcal{D}$, especially for $\mathcal{O}$. 

Since we adopt a wide\&deep structure for our CVR estimator, as shown in Fig. \ref{structure_twin_tower}, the prediction function of our twin tower can be specifically formulated as follows.
\begin{equation}\label{Equation_twin_tower_details}
\small{
	\begin{aligned}
	< \hat{r}_{i,j}, \hat{r}^*_{i,j} > =& < \sigma(l_f),\sigma(l_{cf}) > \\
	=& < \sigma(l^w_f+l^d_f),\sigma(l^w_{cf}+l^d_{cf}) > \\
	=&< \sigma(\phi(\vec{x}^w_{i,j}; \theta^w_f)+\psi(\vec{x}^d_{i,j}; \theta^d,\theta^d_f)),\\ &\sigma(\phi(\vec{x}^w_{i,j}; \theta^w_{cf})+\psi(\vec{x}^d_{i,j}; \theta^d,\theta^d_{cf})) >,
	\end{aligned}
	}
\end{equation}
where $\sigma(*)$ is the \emph{Sigmoid} function, $\phi(\vec{x}, \theta)$ is the \emph{Linear Regression} function with the input embedding $\vec{x}$ and the parameters $\theta$, and $\psi(\vec{x}, \theta)$ is the \emph{Multi-Layer Perceptron (MLP)} function with the input embedding $\vec{x}$ and the parameters $\theta$. In addition, $l_f$ is the factual CVR logit, $l_{cf}$ is the counterfactual CVR logit, $l^w$ (e.g., $l^w_f$ and $l^w_{cf}$) is the CVR logit in the wide part, $l^d$ (e.g., $l^d_f$ and $l^d_{cf}$) is the CVR logit in the deep part, $\vec{x}^w_{i,j}$ is the wide feature embedding vector of the user-item pair $<u_i, v_j>$, $\vec{x}^d_{i,j}$ is the deep feature embedding vector of the user-item pair $<u_i, v_j>$, $\theta^w_f$ is the parameters of $\phi$ for $l^w_f$ in the wide part, $\theta^w_{cf}$ are the parameters of $\phi$ for $l^w_f$ in the wide part, $\theta^d$ are the common parameters of $\psi$ in the deep part, $\theta^d_f$ are the specific parameters of $\psi$ for $l^d_f$ in the deep part, and $\theta^d_{cf}$ are the specific parameters of $\psi$ for $l^d_{cf}$ in the deep part. For readability, we carefully mark these notations in the structure of the twin tower (see Fig. \ref{structure_twin_tower}).


\subsection{Self-Normalization}
To reduce the variance of IPW-based approaches, the Self-Normalized Inverse Propensity Scoring (SNIPS) estimator is widely used in the literature \cite{swaminathan2015self,schnabel2016recommendations}. We also adopt this self-normalization strategy to reduce the variance of our DCMT estimator. Finally, the inverse propensity weights $\frac{1}{\hat{o}_{i,j}}$ and $\frac{1}{(1-\hat{o}_{i,j})}$ in Eq. (\ref{Equation_cvr_loss_our_method_full}) will be replaced with the following two self-normalized weights. 
\begin{equation}\label{self-normalized_weights}
\small{
	\begin{aligned}
	& \frac{\frac{1}{\hat{o}_{i,j}}}{\sum \limits_{(i,j) \in \mathcal{O}} \frac{1}{\hat{o}_{i,j}}}~~~\textrm{and}~~~ \frac{\frac{1}{1-\hat{o}_{i,j}}}{\sum \limits_{(i,j) \in \mathcal{N}^*} \frac{1}{1-\hat{o}_{i,j}}}.
	\end{aligned}
	}
\end{equation}
Also, to avoid the Nan loss in the training process, we need to clip the value range of $\hat{o}_{i,j}$ from the default range, i.e., $[0, 1]$, to $(0, 1)$.

\subsection{Training Loss}
As mentioned in Section \ref{Section_Introduction}, we use the two auxiliary tasks, i.e., \emph{CTR Task} and \emph{CTCVR Task}, to help to obtain the predictions of the target \emph{CVR Task}. Therefore, the training loss of our DCMT framework contains three parts, i.e., the loss of \emph{CTR Task}, the loss of \emph{CVR Task}, and the loss of \emph{CTCVR Task}, as follows:
\begin{equation}\label{Equation_Total_Loss}
\small{
\begin{aligned}
&L(\theta)= \mathcal{E}^{\textrm{CTR}} + w^{cvr}\mathcal{E}^{\textrm{DCMT}} + w^{ctcvr}\mathcal{E}^{\textrm{CTCVR}} + \lambda_2\| \theta \|_{F}^2,
\end{aligned}
}
\end{equation}
where $\mathcal{E}^{\textrm{DCMT}}$ is the CVR loss of our DCMT framework (see Eq. (\ref{Equation_cvr_loss_our_method_full})), $\theta$ denotes all network weights in our DCMT framework, $\| \theta \|_{F}^2$ is the regularizer, $\lambda_2$ is a hyper-parameter that controls the importance of the regularizer. In addition, $w^{cvr}$, $w^{ctcvr}$ are loss weights of $\mathcal{E}^{\textrm{DCMT}}$, $\mathcal{E}^{\textrm{CTCVR}}$, which are set to $1$ in this paper, respectively. The losses of the CTR task and the CTCVR task can be represented as follows.

\begin{equation}\label{Equation_ctr_ctcvr_loss}
\small{
	\begin{aligned}
	\mathcal{E}^{\textrm{CTR}} = \frac{1}{|\mathcal{D}|} \sum \limits_{(i,j) \in \mathcal{D}} e(o_{i,j},\hat{o}_{i,j}),\\
	\mathcal{E}^{\textrm{CTCVR}} = \frac{1}{|\mathcal{D}|} \sum \limits_{(i,j) \in \mathcal{D}} e(r_{i,j},\hat{t}_{i,j}).
	\end{aligned}
	}
\end{equation}

%% file: sections/experiment.tex
\begin{table}[t]
	\begin{center}
		\caption{Experimental datasets} \label{Datasets}
		\vspace{-2mm}
		\resizebox{0.48\textwidth}{!}{
		\begin{tabular}{c@{  }|@{  }c@{  }|c@{  }|@{  }c@{  }|@{  }c@{  }|@{  }c@{  }|@{  }c@{  }|@{  }c@{  }}
			\hline
			\multicolumn{2}{c@{   }|@{}}{\textbf{Dataset}}&\textbf{\#User}&\textbf{\#Item}&\textbf{Split}&\textbf{\#Exposure}&\textbf{\#Click}&\makecell{\textbf{\#Conv-}\\\textbf{version}}\\
			\hline
			\multirow{10}{*}{\makecell{\textbf{Public}\\\textbf{(offline test)}}}&\multirow{2}{*}{\textbf{Ali-CCP}} &\multirow{2}{*}{0.4M} &\multirow{2}{*}{4.3M} & Train &42.3M & 1.6M &9,K\\
			\cline{5-8}
			&& & & Test& 43M & 1.7M & 9.4K\\
			\cline{2-8}
			&\multirow{2}{*}{\textbf{\makecell{AE-ES}}} &\multirow{2}{*}{0.6M} &\multirow{2}{*}{1.4M} & Train & 22.3M  & 0.57M & 12.9K\\
			\cline{5-8}
			& & & & Test & 9.3M & 0.27M & 6.1K\\
			\cline{2-8}
			&\multirow{2}{*}{\textbf{\makecell{AE-FR}}} &\multirow{2}{*}{0.57M} &\multirow{2}{*}{1.2M}  &Train & 18.2M  & 0.34M & 9K\\
			\cline{5-8}
			& & & &Test & 8.8M & 0.2M & 5.3K\\
			\cline{2-8}
			&\multirow{2}{*}{\textbf{\makecell{AE-NL}}}  &\multirow{2}{*}{0.37M} &\multirow{2}{*}{0.81M}  &Train &12.2M & 0.25M & 8.9K\\
			\cline{5-8}
			&& &&Test  & 5.6M & 0.14M & 4.9K\\
			\cline{2-8}
			&\multirow{2}{*}{\textbf{\makecell{AE-US}}} &\multirow{2}{*}{0.5M}&\multirow{2}{*}{1.3M}&Train & 20M & 0.29M & 7K\\
			\cline{5-8}
			&& & &Test & 7.5M  & 0.16M & 3.9K\\
			\hline
			\multirow{2}{*}{\makecell{\textbf{Industrial}\\\textbf{(online test)}}}&\multirow{2}{*}{\textbf{\makecell{Alipay \\ Search}}} &\multirow{2}{*}{73M}&\multirow{2}{*}{531K}&Train & 665M  & 118M & 88M\\
			\cline{5-8}
			&& & &Test & 162M  & 29M & 22M\\
			\hline
		\end{tabular}
		}
	\end{center}
\end{table}
\begin{table*}[t]
	\begin{center}
		\caption{The comparison of the baselines and our methods} \label{Baselines}
		\vspace{-2mm}
		\resizebox{0.8\textwidth}{!}{
			\begin{tabular}{c@{   }|@{   }c@{   }|@{   }c@{   }|@{   }c@{   }|@{   }c}
				\hline
				\multicolumn{3}{c@{   }|@{   }}{\textbf{Model}}& \makecell{\textbf{Structure}} & \textbf{Main Ideas}\\
				\hline
				\hline
				\multirow{15}{*}{Baselines}&\multirow{1}{*}{\makecell{Parallel MTL Baselines}}& \textbf{ESMM} \cite{ma2018entire}&Shared bottom & \makecell{Feature representation\\ transfer learning}\\
				\cline{2-5}
				&\multirow{8}{*}{\makecell{Multi-gate MTL baselines}}&\textbf{Cross Stitch} \cite{misra2016cross}& Cross-stitch unit& Activation combination \\ 
				\cline{3-5}
				&&\textbf{MMOE} \cite{ma2018modeling}&Gated mixture-of-experts&\makecell{Trade-offs between \\task-specific objectives \\and inter-task relationships}\\ 
				\cline{3-5}
				&&\textbf{PLE} \cite{tang2020progressive}& \makecell{Customized gates \& \\local experts \& shared experts}&\makecell{ Customized sharing \\(avoiding negative transfer)}\\ 
				\cline{3-5}
				&& \textbf{AITM} \cite{xi2021modeling}&\makecell{Shared bottom \& \\inter-task transfer}&Adaptive information transfer\\
				\cline{2-5}
				&\multirow{3}{*}{\makecell{Causal baselines}}& \textbf{ESCM$^2$-IPW} \cite{wang2022escm}&\makecell{Two towers \\(CTR+CVR)}& \makecell{Propensity-based debiasing}\\
				\cline{3-5}
				&& \textbf{ESCM$^2$-DR} \cite{wang2022escm}&\makecell{Three towers \\(CTR+CVR+Imputation)}&\makecell{Propensity-based debiasing\\ \& doubly robust estimation}\\
				\hline
				\hline
				\multirow{4}{*}{\makecell{Our\\ methods}}&\multirow{2}{*}{\makecell{Simplified versions}}&\makecell{\textbf{DCMT\_PD}}&\makecell{CTR tower + the twin CVR tower} & Propensity-based debiasing over $\mathcal{D}$\\ \cline{3-5}
				&&\makecell{\textbf{DCMT\_CF}}&\makecell{CTR tower + the twin CVR tower} & \makecell{Counterfactual mechanism}\\ 
				\cline{2-5}
				&Completed version&\textbf{DCMT}&\makecell{CTR tower + the twin CVR tower}  & \makecell{Propensity-based debiasing\&\\ counterfactual mechanism }\\  
				\hline
			\end{tabular}
		}
	\end{center}
	\vspace{-5mm}
\end{table*}
\section{Experiments and Analysis} \label{Experiments&Analysis}
We conduct extensive experiments on real-world benchmark datasets and online systems to answer the following key questions: 
\begin{itemize}[leftmargin=*]
\item \textbf{Q1} How does our DCMT model perform when compared with the state-of-the-art (SOTA) models on offline datasets (see Result 1)? 
\item \textbf{Q2} How much does the counterfactual mechanism contribute to performance improvement on offline datasets (see Result 2)? 
\item \textbf{Q3} How does our DCMT model perform when compared with the base model in online environments (see Result 3)? 
\item \textbf{Q4} How do the hyper-parameters affect the performance of our DCMT model (see Result 4)? 
\end{itemize}

\subsection{Experimental Settings}
We conduct extensive experiments on both the offline datasets collected from real-world e-commerce \& express systems, and online searching environments.

\begin{table*}[t]
	\begin{center}
		\caption{The offline experimental results (AUC) in different public datasets for \emph{CVR Task} and \emph{CTCVR Task} (the best-performing baselines with results marked with \textbf{*} while our best-performing models with results marked with black body)} \label{ExperimentalResults}
		\vspace{-2mm}
		\resizebox{0.99\textwidth}{!}{
			\begin{tabular}{@{}c@{}||@{}c@{}||@{}c@{}|@{}c@{}|@{}c@{}|@{}c@{}||@{}c@{}|@{}c@{}||@{}c@{}|@{}c@{}|@{}c@{}||@{}c@{}}
				\hline
				\multirow{3}{*}{\textbf{Dataset}}& \multicolumn{1}{c@{}||@{}}{\textbf{\makecell{Parallel MTL\\ Baselines}}} & \multicolumn{4}{c@{}||@{}}{\makecell{\textbf{Multi-gate MTL Baselines}}}&\multicolumn{2}{c@{}||@{}}{\makecell{\textbf{Causal Baselines}}}&\multicolumn{3}{c@{}||@{}}{\makecell{\textbf{Our Models (for ablation study)}}}& \makecell{\textbf{Improvement} \\\textbf{(DCMT} vs. \textbf{best}} \\
				\cline{2-11}
				& ESMM & Cross Stitch & MMOE & PLE  & AITM &ESCM$^2$-IPW& ESCM$^2$-DR & DCMT\_PD & DCMT\_CF & DCMT & \textbf{-performing baselines)}\\
				\cline{2-12}
				&CVR~~CTCVR&CVR~~CTCVR&CVR~~CTCVR&CVR~~CTCVR&CVR~~CTCVR&CVR~~CTCVR&CVR~~CTCVR&CVR~~CTCVR&CVR~~CTCVR&CVR~~CTCVR&CVR~~CTCVR\\
				\hline
				\hline
				Ali-CCP&.6291~.6243*&.5637~.5584&.6041~.5947&.5871~.5599&.6324*~.6121&.6156~.5932&.5914~.6195&.6352~.6156&.6366~.6141&\textbf{.6486}~\textbf{.6341}&\color{red}2.56\%$\uparrow$~1.57\%$\uparrow$\\
				\hline
				\hline
				\makecell{AE-ES}&.6293~.6027&.6855~.6453&.6082~.6211&.5262~.6146&.7109~.7044&.8561~.8688&.8601*~.8681*&.8422~.8329&.8612~.8557&\textbf{.8677}~\textbf{.8817}&\color{red}0.88\%$\uparrow$~1.57\%$\uparrow$ \\
				\hline
				\hline
				\makecell{AE-FR}&.6175~.6039&.7186~.6443&.6290~.0.6451&.5827~.5351&.7316~.6923&.8524*~.8647&.8513~.8668*&.8087~.8512&.8271~.8470&\textbf{.8576}~\textbf{.8756}&\color{red}0.61\%$\uparrow$~1.02\%$\uparrow$ \\
				\hline
				\hline
				\makecell{AE-NL}&.5951~.5329&.7207~.7027&.5050~.7173&.5328~.7011&.5454~.5031&.8323*~.8486*&.7196~.7196&.8212~.8270&.8117~.8143&\textbf{.8331}~\textbf{.8513}&\color{red}0.10\%$\uparrow$~0.32\%$\uparrow$ \\
				\hline
				\hline
				\makecell{AE-US}&.5086~.5881&.6114~.6808&.5395~.6358&.4627~.5793&.5387~.5447&.8385*~.8385*&.7625~.7625&.8334~.8390&.8433~.8404&\textbf{.8484}~\textbf{.8629}&\color{red}1.18\%$\uparrow$~2.91\%$\uparrow$\\
				\hline
			\end{tabular}
		}
	\end{center}
	\vspace{-5mm}
\end{table*}
\subsubsection{Experimental Datasets}
To validate the prediction performance of our DCMT model and baseline models, we choose two real-world benchmark datasets, i.e., Ali-CCP\footnote{Dataset URL: https://tianchi.aliyun.com/dataset/dataDetail?dataId=408} (Alibaba
Click and Conversion Prediction) \cite{ma2018entire} and Ali-Express (AE)\footnote{Dataset URL: https://tianchi.aliyun.com/dataset/dataDetail?dataId=74690} \cite{peng2020improving}. Since Ali-Express includes the real-world traffic logs of the Ali-Express search system from different countries, we choose four suitable sub-datasets from four countries, i.e., Spain (AE-ES), French (AE-FR), Netherlands (AE-NL), and America (AE-US). Both datasets encompass user features, item features, combination features (only in Ali-CCP), context features (only in Ali-CCP), and labels of click and conversion. Also, we validate the performance of our DCMT on our Alipay Search dataset (industrial dataset). The Alipay Search dataset comes from a 10-day offline log of the service searching of our Alipay platform, divided into training (7 days), validation (1 day), and testing (2 days) in chronological order. The Alipay Search dataset includes a random sample of the users' searching historical data on the services provided by our Alipay platform. We treat the click on the detailed pages of services, i.e., the second click, as conversion. To protect information privacy, this industrial dataset has been encrypted and desensitized, which is only used for academic research. For a clear comparison, we list the dataset statistics in Table \ref{Datasets}.

\subsubsection{Parameter Setting} \label{section_parameter_setting}
To ensure a fair comparison, we align the hyper-parameters of our DCMT with those baseline models according to the parameter settings reported in their original papers. For \emph{Embedding Layer} of our DCMT model, we set the embedding dimension for each feature (sparse id feature or dense numerical feature or weighted feature) to $32$ owing to the best performance observed from the experiment (see Fig. \ref{parameter_of_embedding_dimension}). For the deep towers in \emph{CTR Task} and \emph{CVR Task} in Fig. \ref{our_model_structure}, the MLP structure of the layers is [64-64-32] for the AE datasets, [320-200-80] for the Ali-CCP dataset, and [512-256-128] for the industrial dataset (see Fig. \ref{parameter_of_mlp_structure}). For training our DCMT model, we adopt Adam \cite{kingma2014adam} to train the neural networks and set the maximum number of training epochs to $5$. The learning rate is $0.001$, the weight of the counterfactual regularizer $\lambda_1$ is $0.001$, the regularization coefficient $\lambda_2$ is $0.0001$, and the batch size is $1,024$.

\subsubsection{Evaluation Metrics}
For the offline test, since the main task for this work is to predict post-click conversion rate, in the experiments, we report the experimental results of \emph{CVR Task} and \emph{CTCVR Task}. The area under the ROC curve (AUC) is adopted as performance metrics in the experiments. All experiments are repeated 5 times and averaged results are reported. Also, for the online A/B test, we mainly focus on the three online business metrics, i.e., PV-CTR (the click-through rate for each page view), PV-CVR (the conversion rate for each page view), and Top-5 PV-CVR (PV-CVR for the top-5 services, a maximum of 5 services can be displayed on one screen). From the perspective of business, enhancing the click-through rate and conversion (the double click) rate for each page view from users can significantly improve the user experience and attract more new users.

\subsubsection{Comparison Methods}
As shown in Table \ref{Baselines}, we compare our DCMT model with seven baseline models in three groups, i.e., (1) parallel MTL baselines, (2) multi-gate MTL baselines, and (3) causal baselines. All seven baselines are representative and/or state-of-the-art approaches for each group. To make an ablation study, we implement other two simplified versions as the variants of the completed DCMT, i.e., DCMT\_PD (only considering the propensity-based debiasing in the factual exposure space $\mathcal{D}$) and DCMT\_CF (only adopting the counterfactual mechanism). For a clear comparison, in Table \ref{Baselines}, we list the detailed structures, task relationships, and main ideas of all the models implemented in the experiments.
\begin{table}[t]
	\begin{center}
		\caption{The online experimental results of the A/B test (a week in August 2022) on the Alipay Search system. $\uparrow$ represents the increase compared with the base model MMOE, $\downarrow$ represents the decrease, the pink background colour represents the increase with 95\% confidence intervals, and the green background colour represents the decrease with 95\% confidence intervals.} \label{online_test_results}
		\vspace{-4mm}
		\resizebox{0.49\textwidth}{!}{
		\begin{tabular}{@{}c@{}|@{}c@{}|@{}c@{}|@{}c@{}|@{}c@{}|@{}c@{}|@{}c@{}|@{}c@{}|@{}c@{}|@{}c@{}}
			\hline
			Metric& Model &Day1&Day2&Day3&Day4&Day5&Day6&Day7&Overall\\
			\hline
			\multirow{3}{*}{PV-CTR}&ESCM$^2$-IPW&\color{red}0.47\%$\uparrow$&\fcolorbox{green}{green}{1.00\%$\downarrow$}&\color{red}0.14\%$\uparrow$&\fcolorbox{green}{green}{0.95\%$\downarrow$}&0.43\%$\downarrow$&0.23\%$\downarrow$&\color{red}0.28\%$\uparrow$&\fcolorbox{green}{green}{0.32\%$\downarrow$}\\
			\cline{2-10}
			&ESCM$^2$-DR&0.23\%$\downarrow$&0.43\%$\downarrow$&\color{red}0.13\%$\uparrow$&\fcolorbox{green}{green}{0.61\%$\downarrow$}&0.06\%$\downarrow$&0.39\%$\downarrow$&0.03\%$\downarrow$&0.24\%$\downarrow$\\
			\cline{2-10}
			&DCMT&0.32\%$\downarrow$&0.44\%$\downarrow$&\fcolorbox{pink}{pink}{0.91\%$\uparrow$}&\color{red}0.36\%$\uparrow$&\fcolorbox{pink}{pink}{0.99\%$\uparrow$}&\fcolorbox{pink}{pink}{0.67\%$\uparrow$}&\fcolorbox{pink}{pink}{1.26\%$\uparrow$}&\fcolorbox{pink}{pink}{0.49\%$\uparrow$}\\
			\hline
			\multirow{3}{*}{PV-CVR}&ESCM$^2$-IPW&\color{red}0.43\%$\uparrow$&\fcolorbox{green}{green}{1.15\%$\downarrow$}&\color{red}0.25\%$\uparrow$&\fcolorbox{green}{green}{0.91\%$\downarrow$}&0.42\%$\downarrow$&0.00\%&\fcolorbox{pink}{pink}{0.68\%$\uparrow$}&0.22\%$\downarrow$\\
			\cline{2-10}
			&ESCM$^2$-DR&0.04\%$\downarrow$&0.28\%$\downarrow$&\color{red}0.44\%$\uparrow$&0.57\%$\downarrow$&\color{red}0.26\%$\uparrow$&0.31\%$\downarrow$&0.21\%$\downarrow$&0.02\%$\downarrow$\\
			\cline{2-10}
			&DCMT&\color{red}0.51\%$\uparrow$&0.42\%$\downarrow$&\fcolorbox{pink}{pink}{1.49\%$\uparrow$}&\color{red}0.53\%$\uparrow$&\fcolorbox{pink}{pink}{0.89\%$\uparrow$}&\fcolorbox{pink}{pink}{0.71\%$\uparrow$}&\fcolorbox{pink}{pink}{1.42\%$\uparrow$}&\fcolorbox{pink}{pink}{0.75\%$\uparrow$}\\
			\hline
			\multirow{3}{*}{\makecell{Top-5\\ PV-CVR}}&ESCM$^2$-IPW&\color{red}0.38\%$\uparrow$&\fcolorbox{green}{green}{1.09\%$\downarrow$}&\color{red}0.26\%$\uparrow$&\fcolorbox{green}{green}{0.85\%$\downarrow$}&0.41\%$\downarrow$&\color{red}0.01\%$\uparrow$&\fcolorbox{pink}{pink}{0.66\%$\uparrow$}&0.21\%$\downarrow$\\
			\cline{2-10}
			&ESCM$^2$-DR&0.07\%$\downarrow$&0.29\%$\downarrow$&\color{red}0.47\%$\uparrow$&0.48\%$\downarrow$&\color{red}0.26\%$\uparrow$&0.35\%$\downarrow$&0.20\%$\downarrow$&0.02\%$\downarrow$\\
			\cline{2-10}
			&DCMT&\color{red}0.47\%$\uparrow$&0.67\%$\downarrow$&\fcolorbox{pink}{pink}{1.59\%$\uparrow$}&\color{red}0.58\%$\uparrow$&\fcolorbox{pink}{pink}{0.88\%$\uparrow$}&\fcolorbox{pink}{pink}{0.74\%$\uparrow$}&\fcolorbox{pink}{pink}{1.51\%$\uparrow$}&\fcolorbox{pink}{pink}{0.66\%$\uparrow$}\\
			\hline
		\end{tabular}
		}
	\end{center}
	\vspace{-5mm}
\end{table}
\begin{figure*}[t]
 	\begin{center}
 	 	\subfigure[The CVR prediction distribution of the base model (MMOE) over $\mathcal{D}$ (the average CVR prediction is $0.320$ that is close to $\beta$)]{
 		\label{Base_prediction_distribution}
 		\vspace{-2mm}
 		\includegraphics[width=0.38\textwidth]{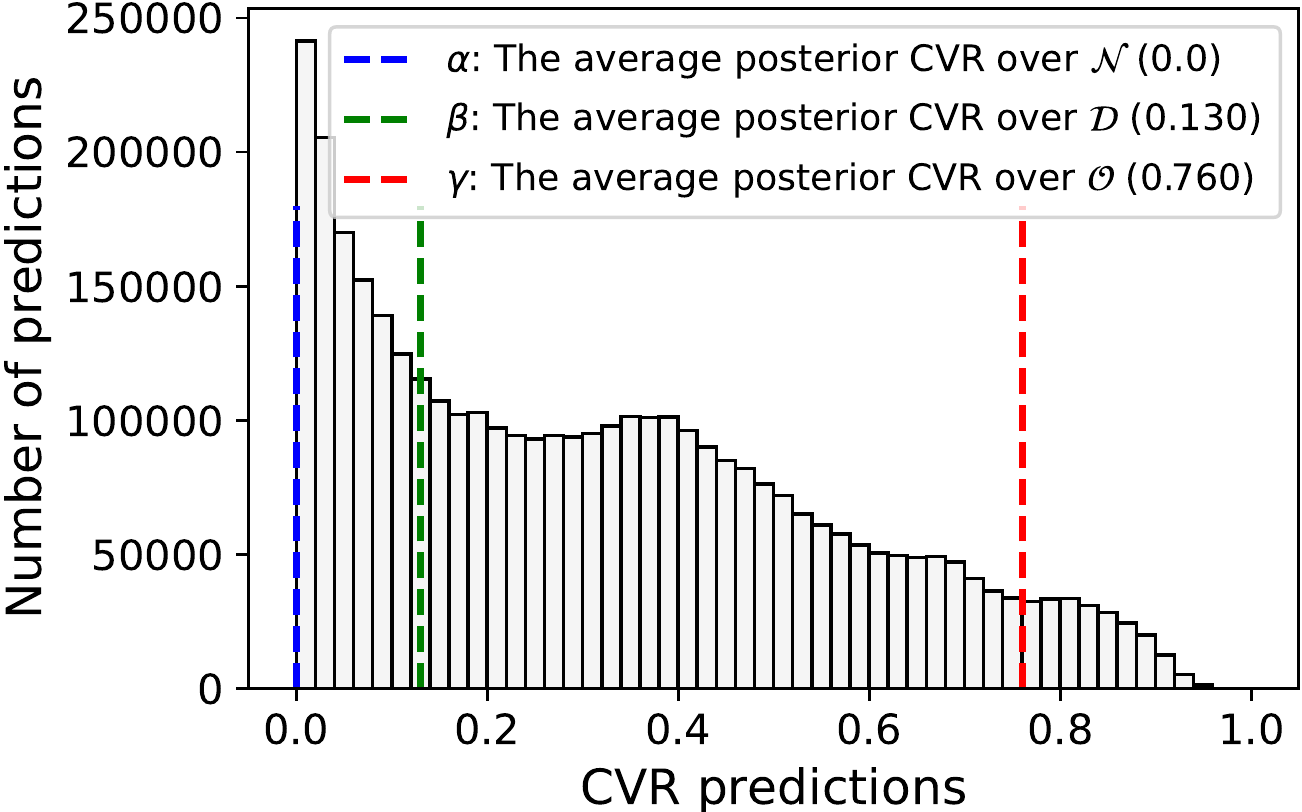}}
 		\hspace{2mm}
 		 \subfigure[The CVR prediction distribution of ESCM$^2$-IPW over $\mathcal{D}$ (the average CVR prediction is $0.676$ that is close to $\gamma$)]{
 		\label{IPW_prediction_distribution}
 		\vspace{-2mm}
 		\includegraphics[width=0.38\textwidth]{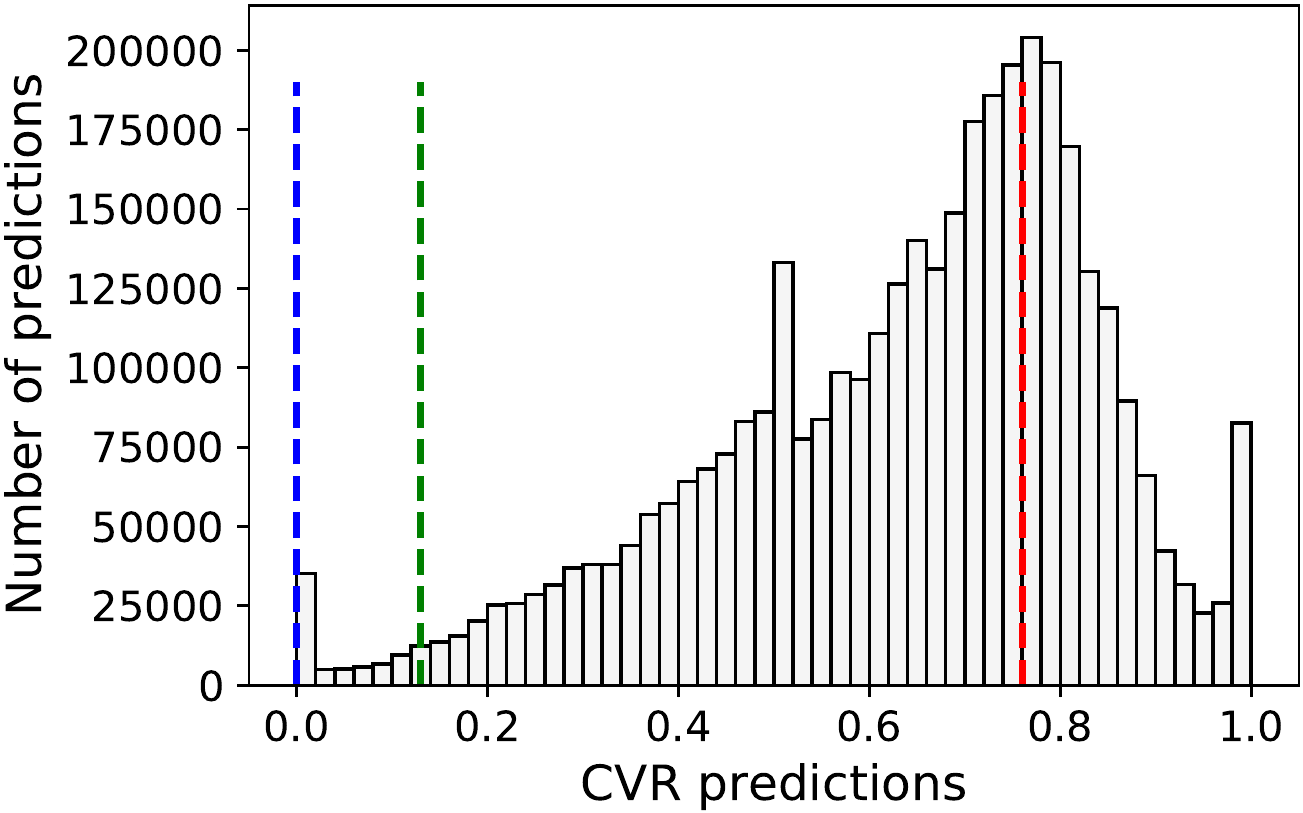}}
 		 \subfigure[The CVR prediction distribution of ESCM$^2$-DR over $\mathcal{D}$ (the average CVR prediction is $0.637$ that is close to $\gamma$)]{
 		\label{DR_prediction_distribution}
 		\vspace{-2mm}
 		\includegraphics[width=0.38\textwidth]{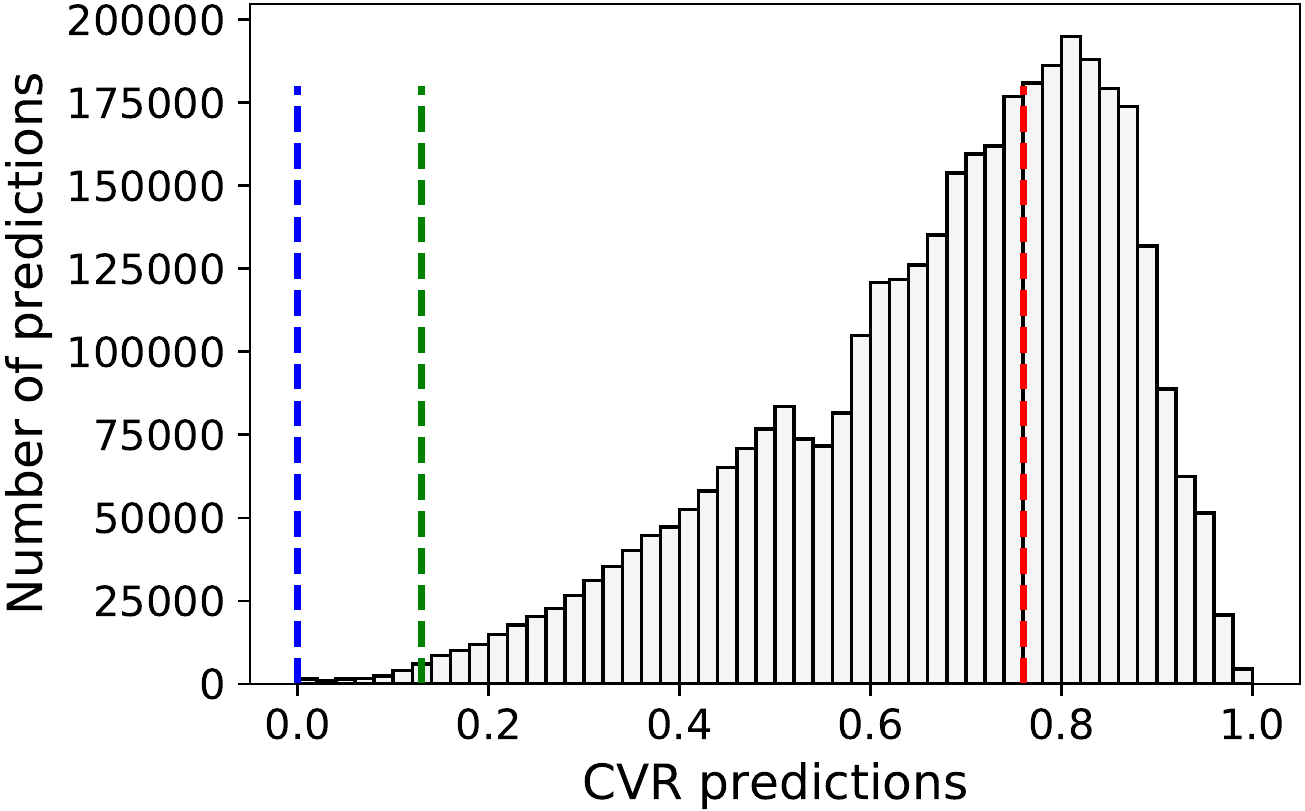}}
 		\hspace{2mm}
 		\subfigure[The CVR prediction distribution of DCMT over $\mathcal{D}$ (the average CVR prediction is $0.343$ that is close to $\beta$)]{
 		\label{DCMT_prediction_distribution}
 		\vspace{-2mm}
 		\includegraphics[width=0.38\textwidth]{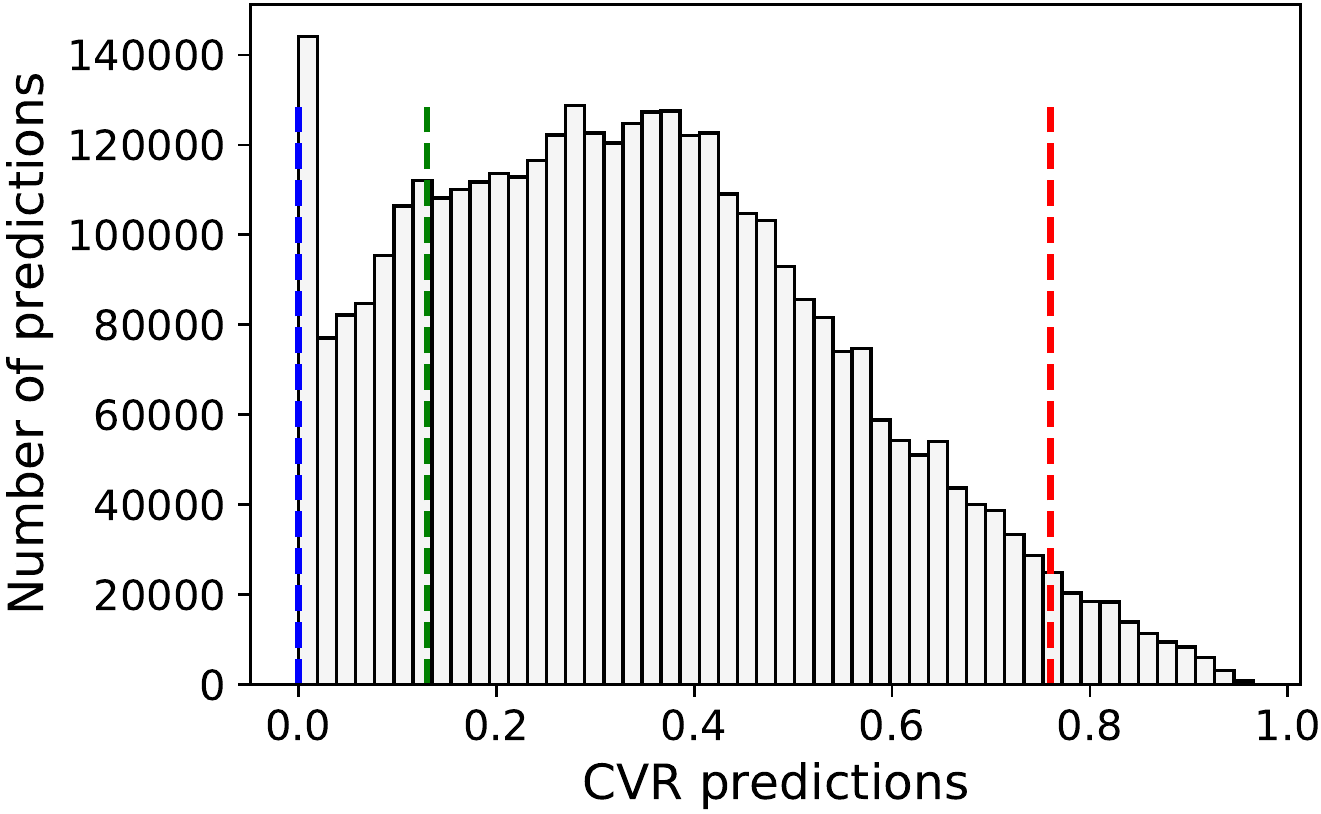}}
 		\vspace{-2mm}
 		\caption{The online prediction distribution of CVR over $\mathcal{D}$.}
 	\label{CVR_prediction_distribution}
 	\end{center}
 	\vspace{-5mm}
 \end{figure*}
\subsection{Performance Comparison and Analysis}
\subsubsection{Offline test on the public datasets (\textbf{Results 1 and 2})}

\nosection{\textbf{Result 1: Offline performance comparison (for Q1)}}
Table \ref{ExperimentalResults} shows the experimental results in term of AUC in different datasets for \emph{CVR Task} and \emph{CTCVR Task}, respectively.
To answer \textbf{Q1}, we compare the performance of our DCMT with those of the seven baseline models. As indicated in Table \ref{ExperimentalResults}, our DCMT outperforms the best-performing baselines by an average improvement of $1.07\%$ in term of CVR AUC (the main task). In particular, our DCMT improves the best-performing baselines (with results marked by * in Table \ref{ExperimentalResults}) by $2.56\%$ on the Ali-CCP dataset, $0.88\%$ on the AE-ES dataset, $0.61\%$ on AE-FR, $0.10\%$ on AE-NL, and $1.18\%$ on AE-US. This is because our DCMT can debias the selection bias over $\mathcal{D}$, demonstrating the superiority in terms of easing selection bias and data sparsity.
 \begin{figure*}[t]
	\begin{center}
	\subfigure[The impact of feature embedding dimension (AE-ES)]{
 		\label{parameter_of_embedding_dimension}
 		\vspace{-2mm}
 		\includegraphics[width=0.45\textwidth]{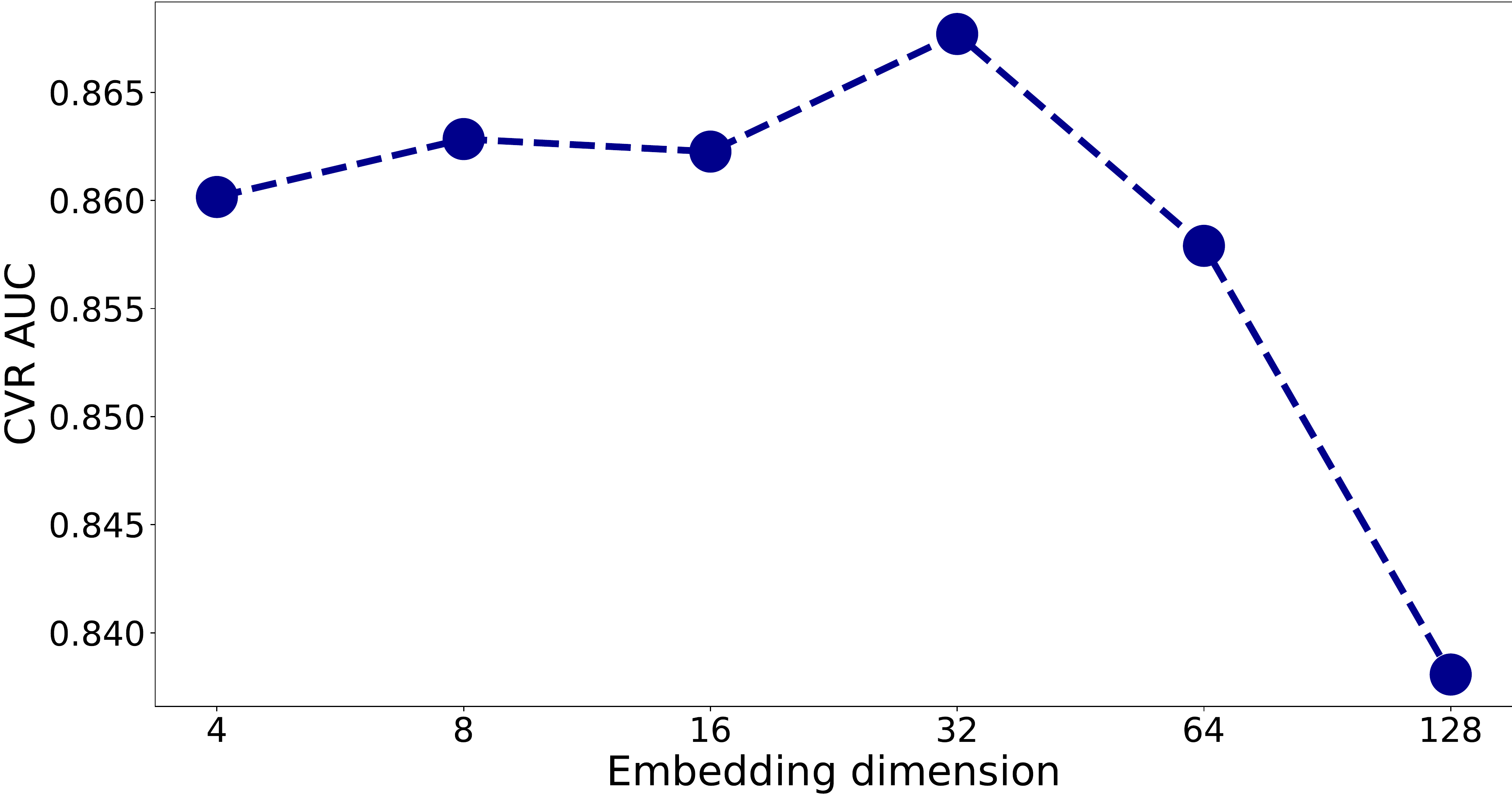}}
 	\subfigure[The impact of MLP structure (AE-ES)]{
 		\label{parameter_of_mlp_structure}
 		\vspace{-2mm}
 		\includegraphics[width=0.48\textwidth]{images/impact_of_mlp_structure.pdf}}
 	\subfigure[The impact of counterfactual regularizer weight $\lambda_1$ (AE-ES)]{
 		\label{parameter_of_lambda1}
 		\vspace{-2mm}
 		\includegraphics[width=0.48\textwidth]{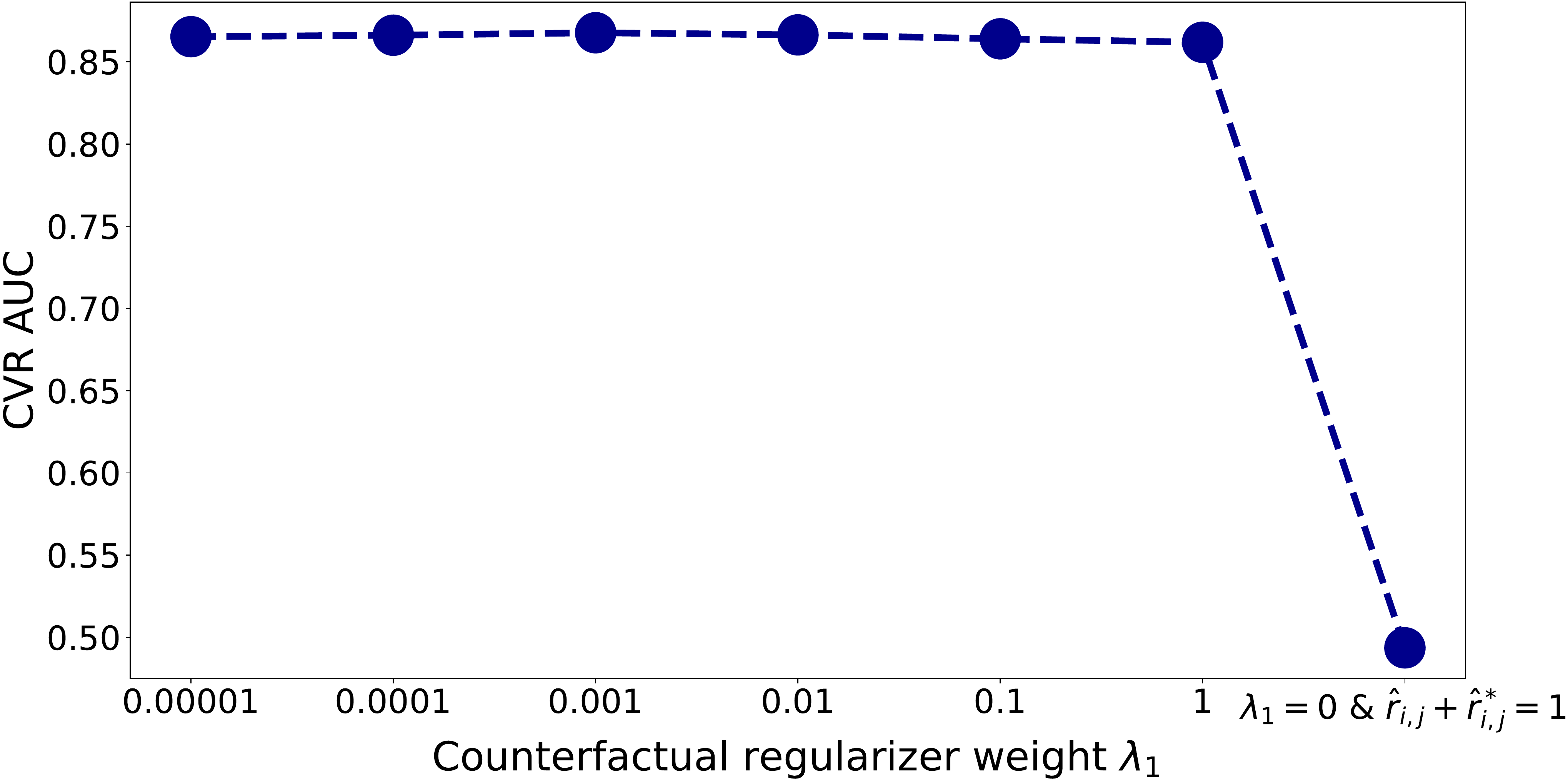}}
	\subfigure[Prediction ranges of CVR ($\hat{r}_{i,j}+\hat{r}^*_{i,j}=1$, AE-ES)]{
 		\label{parameter_of_prediction_ranges}
 		\vspace{-2mm}
 		\includegraphics[width=0.45\textwidth]{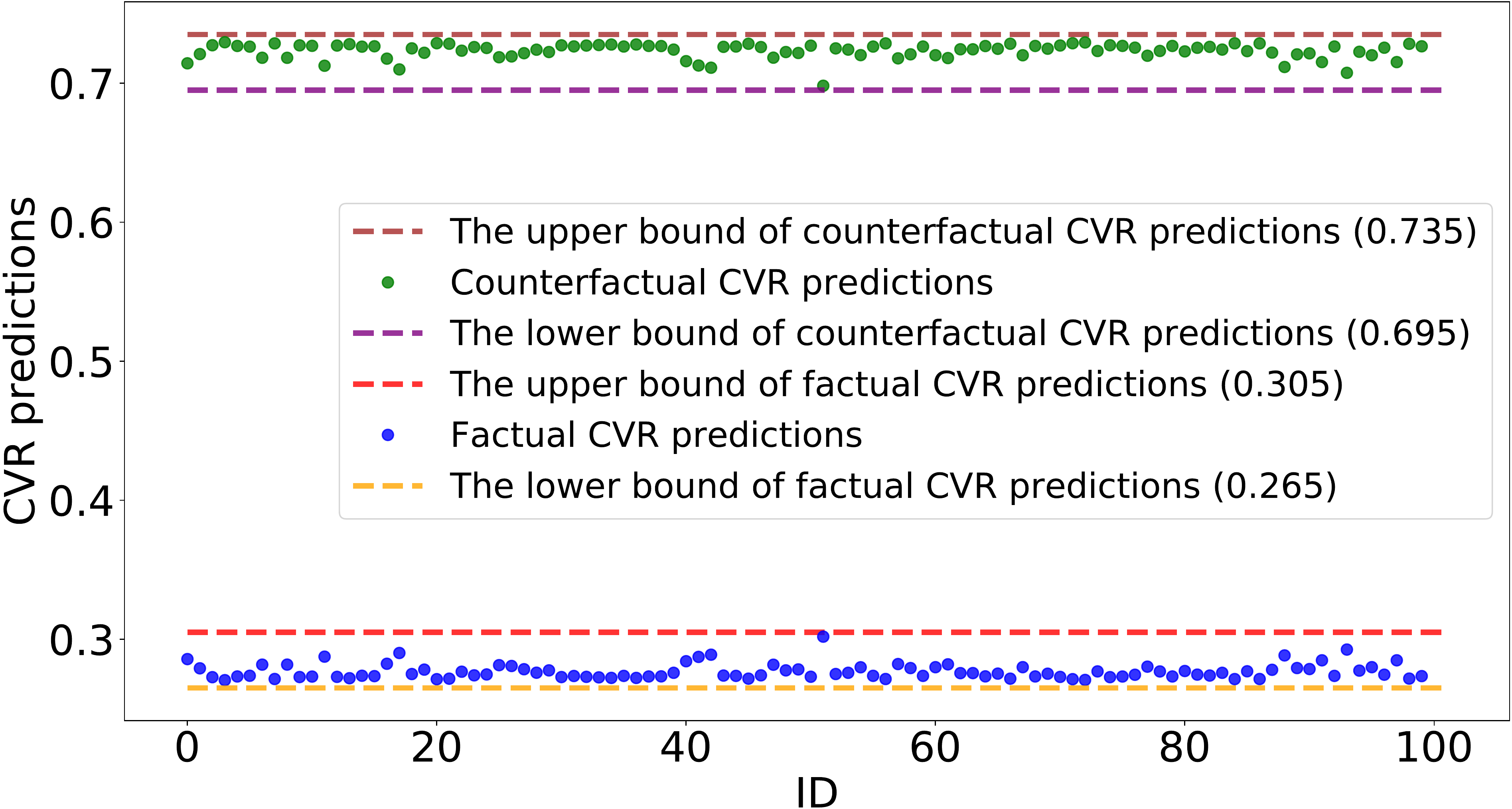}}
		\caption{Impact of Hyper-parameters.}
		\label{impact_of_hp}
	\end{center}
	\vspace{-5mm}
\end{figure*}

\nosection{\textbf{Result 2: Ablation study (for Q2)}}
To answer \textbf{Q2}, we implement two variants of our DCMT, i.e., DCMT\_PD (only considering the propensity-based debiasing in $\mathcal{D}$) and DCMT\_CF (only adopting the counterfactual mechanism) to demonstrate the detailed contributions of the two main components of our DCMT framework. 

On the one hand, as we can see from Table \ref{ExperimentalResults}, when we only consider the propensity-based debiasing (DCMT\_PD) in $\mathcal{D}$, our completed DCMT improves DCMT\_PD by an average of $2.89\%$ for CVR AUC. This result indicates that our counterfactual mechanism can further improve the prediction performance and effectively alleviate the problems of selection bias and data sparsity.

On the other hand, compared with DCMT\_CF, our completed DCMT model achieves an average improvement of $1.91\%$ (according to the results in Table \ref{ExperimentalResults}). This means that the propensity-based debiasing (DCMT\_PD) in $\mathcal{D}$ also plays a very important role in our DCMT.
\subsubsection{Online A/B test on the Alipay platform (\textbf{Result 3})} \label{online_experimental_results}

\nosection{\textbf{Result 3-1: Online performance comparison (for Q3)}} To answer \textbf{Q3}, in addition to the above-mentioned offline experiments, we also conduct extensive online A/B experiments on the Alipay Search platform to further validate the performance of our DCMT framework. In the online setting, an MMOE-based model is the base model (the current running model), and we also choose two state-of-the-art models, i.e., ESCM$^2$-IPW, and ESCM$^2$-DR, as the online baselines. Specifically, we first implement these baselines and our DCMT with our TensorFlow-based machine learning framework and deploy them on our inference platform. We then randomly assign the same number of users into the corresponding four buckets, i.e., MMOE (the base model), ESCM$^2$-IPW, ESCM$^2$-DR, and DCMT, via using our online A/B testing platform. Finally, we observe each model's online performance in the respective bucket of users on the A/B testing platform. From the perspective of business, we mainly focus on the three online metrics, i.e., PV-CTR, PV-CVR, and Top-5 PV-CVR. Finally, we report the online A/B testing results of a week (in August 2022) in Table \ref{online_test_results}. 

This online experiment lasts 7 days and covers around 4.1 million unique visitors (UVs) and around 8.9 million page views (PVs) for each experiment bucket. Overall, our DCMT framework improves the base model (MMOE) by 0.49\% (for PV-CTR) with 95\% confidence intervals, 0.75\% (for PV-CVR) with 95\% confidence intervals, and 0.66\% (for Top-5 PV-CVR) with 95\% confidence intervals. Also, as we can see from Table \ref{online_test_results}, our DCMT can consistently outperform both ESCM$^2$-IPW and ESCM$^2$-DR in terms of PV-CTR, PV-CVR, and Top-5 PV-CVR.

\nosection{\textbf{Result 3-2: Online prediction comparison (for Q3)}} To clearly compare the CVR prediction performance of our DCMT with those of the two online baselines, i.e., ESCM$^2$-IPW, and ESCM$^2$-DR, we collect the online predictions of the four online methods over the infer space $\mathcal{D}$ from the log of Day 1 on our online A/B testing platform and draw their distributions in Figure \ref{CVR_prediction_distribution}. Note that in the online environment, we cannot obtain the ground-truth distribution of CVR predictions. Thus, to compare the CVR prediction performance of the four online methods, we mark the average posterior CVR over $\mathcal{D}$, $\mathcal{O}$, and $\mathcal{N}$ respectively in Figure \ref{CVR_prediction_distribution}. As we can see from Figure  \ref{CVR_prediction_distribution}, the average CVR predictions of ESCM$^2$-IPW ($0.676$), and ESCM$^2$-DR ($0.637$) over $\mathcal{D}$ are close to the average posterior CVR over $\mathcal{O}$ ($0.760$) and away from the average posterior CVR over $\mathcal{D}$ ($0.130$). This means that only debiasing over $\mathcal{O}$ (ESCM$^2$-IPW and ESCM$^2$-DR) cannot infer very well over $\mathcal{D}$. Although the average CVR prediction of the base model (MMOE) is close to the average posterior CVR over $\mathcal{D}$ ($0.130$), there are many predictions concentrating between the average posterior CVR over $\mathcal{N}$ ($0.0$) and the average posterior CVR over $\mathcal{N}$ ($0.130$). In contrast, the prediction majority of our DCMT concentrates between the average posterior CVR over $\mathcal{D}$ ($0.130$) and the average posterior CVR over $\mathcal{O}$ ($0.760$), which is significantly better than those of the other three online baselines. This result demonstrates that our DCMT can effectively debias the selection bias in $\mathcal{D}$.
%
\subsection{Impact of Hyper-parameters (Result 4)}\label{section_impact_of_hp}
To answer \textbf{Q4}, in this section, we analyse the impacts of the dimension of feature embedding, the structure of MLP in our DCMT, and the weight of the counterfactual regularizer $\lambda_1$. Due to space limitations, we only report the offline experimental results of the CVR task on the AE-ES dataset.

\nosection{Impact of feature embedding dimension}
To study the impact of feature embedding dimension on our DCMT framework, we choose different dimensions, i.e., $\{4, 8, 16, 32, 64, 128\}$, for performance comparison. The results of \emph{CVR Task} on AE-ES are reported in Fig. \ref{parameter_of_embedding_dimension}. As observed from Fig. \ref{parameter_of_embedding_dimension}, when the embedding dimension is $16$, our DCMT model achieves the best performance. From $16$ to $128$, the performance decreases with the embedding dimension because a large dimension may make our DCMT framework over-fitting to the training samples.

 \nosection{Impact of MLP structure}
To study the impact of MLP structure on our DCMT framework, we attempt to set the depth of MLP in the deep towers (see Fig. \ref{our_model_structure}) from $1$ to $6$ with different numbers of units. Due to space limitations, we only report the best-performing structure for each depth, e.g., $[128]$ for $d_{mlp}=1$ and $[64-64]$ for $d_{mlp}=2$. Similar to the above experiment of embedding dimension, we only report the AUC results of \emph{CVR Task} on AE-ES in Fig. \ref{parameter_of_mlp_structure}. As we can see from Fig. \ref{parameter_of_mlp_structure}, when the structure of MLP is $[64-64-32]$ ($d_{mlp}=3$), our DCMT can achieve the best performance. In general, the performance of our DCMT increases with the depth of MLP from $1$ to $3$. After that, the performance will decrease mainly because a complex MLP structure may lead to an over-fitting problem. Similarly, we also choose the best-performing structures for the Ali-CCP dataset ($[320-200-80]$) and the Alipay Search dataset ($[512-256-128]$).

 \nosection{Impact of counterfactual regularizer weight $\lambda_1$}
To study the impact of counterfactual regularizer weight $\lambda_1$ in our DCMT framework, we choose different weights, i.e., $\{0.00001, 0.0001, 0.001, 0.01, 0.1, 1\}$, for the counterfactual regularizer $L$ in Eq. (\ref{Equation_cvr_loss_our_method_full}). In addition to the soft constraint, we also conduct an experiment to validate the performance of the hard constraint, i.e., we force $\hat{r}_{i,j} + \hat{r}^*_{i,j} =1$. Similarly, we only report the experimental results on the AE-ES dataset in Fig. \ref{parameter_of_lambda1}. 
 In general, the performance of our DCMT increases from $0.00001$ to $0.001$ (the best-performing weight). After that, the performance will decrease mainly because a stronger constraint may make the training process of our CVR estimation difficult to minimize the main loss, i.e., the loss between CVR labels and CVR predictions. Also, the CVR AUC of the hard constraint, i.e., $\lambda_1=0$ and $\hat{r}_{i,j} + \hat{r}^*_{i,j} =1$, is significantly worse than those of the soft constraints in Fig. \ref{parameter_of_lambda1}. To study this issue, we randomly choose 100 samples in the AE-ES dataset, and report their factual CVR predictions and counterfactual CVR predictions in Fig. \ref{parameter_of_prediction_ranges}. The factual CVR predictions and counterfactual CVR predictions of these samples are restricted in small value ranges, i.e., $[0.265, 0.305]$ and $[0.695, 0.735]$. This issue prevents our DCMT to minimize the main loss and thus we did not choose the hard constraint, which has been analysed in Section \ref{section_counterfactual_mechanism} as well.

%% file: sections/relatedwork.tex
\section{Related Work}
\subsection{Multi-Task Learning for CVR Prediction} \label{section_multiple_tasks}
Click-through rate (CTR) and conversion rate (CVR) predictions are two traditional tasks in recommender systems, and some of existing approaches \cite{ma2018entire,pan2019predicting,ouyang2019deep,zhang2020large,liu2020autofis,song2020towards,meng2021general,zhu2021cross,wen2021hierarchically,li2021dual} in the literature attempt to understand users' click and purchase behaviour. In \cite{ma2018entire}, Ma et al. firstly proposed click-through \& conversion rate (CTCVR) as an auxiliary task with CTR task to help improve the prediction accuracy of CVR. Since the scope of this paper is post-click CVR prediction, we only review the highly-related literature on CVR prediction and multiple tasks (including CVR prediction) as follows.

\nosection{CVR Prediction} CVR prediction is a very challenging task because conversed (purchased) samples are very rare in real-world datasets. Recently, with help of deep neural network, there are some related solutions \cite{lu2017practical,pan2019predicting,wen2021hierarchically,bao2020gmcm,wen2020entire,ma2018entire,zhang2020large,su2020attention,xi2021modeling,wen2019multi} focusing on CVR prediction. These approaches adopt some effective feature representation strategies and end-to-end models to study the underlying logic between conversion and input features, e.g., user profiles, item details, exposure position, and users' behaviour. 

\nosection{Multiple Tasks} In \cite{ma2018entire}, to address selection bias and data sparsity problems, the proposed model, i.e., Entire Space Multi-task Model (ESMM), is trained in the entire exposure space. Inevitably, the ESMM may suffer from other bias problems, e.g., exposure bias. The extension works  \cite{pan2019predicting,zhang2020large,wen2020entire,xi2021modeling,wen2021hierarchically} further improve the prediction accuracies of CVR and CTCVR by considering more missing samples and users' feedback actions. Also, some traditional and novel multi-task learning framework, e.g., Cross Stitch \cite{misra2016cross},  MMOE \cite{ma2018modeling}, PLE \cite{tang2020progressive}, and MOSE \cite{qin2020multitask}, are adopted to predict CTR, CVR, and CTCVR, simultaneously. 
\subsection{Causal Inference for Recommendation} \label{section_causal_recommendation}
Causal inference is to study the independent and actual effect of a particular component in a system \cite{pearl2009causal}. Causal inference has been utilized in some recommender systems, aiming to accurately infer users' preferences for items. Early studies \cite{liang2016causal,joachims2017unbiased,ai2018unbiased} mainly focus on debiasing implicit feedback, e.g., position bias \cite{craswell2008experimental}. Next, some researchers start to study the recommendation fairness \cite{morik2020controlling} and other biases, e.g., exposure bias \cite{liang2016modeling}, popularity bias \cite{abdollahpouri2019managing}, in the training samples. Also, there is a novel work, i.e., counterfactual recommendation \cite{wang2021clicks,fang2022interpreting,tang2022debiased}, which adopts counterfactual inference to mitigate the clickbait problem in recommender systems. To further debias the selection bias in the training samples, Inverse Propensity Scoring Weighting (IPW) \cite{rosenbaum1983central} is applied to regenerate the original samples based on a re-weighting strategy. With the help of multi-task learning, recently, some IPW-based debiasing approaches, e.g., Multi-IPW \cite{zhang2020large} and ESCM$^2$-IPW \cite{wang2022escm}, are proposed for CVR estimation. 

%% file: sections/conclusion.tex
\section{Conclusion and Future Work}
In this paper, we have proposed a \textbf{D}irect entire-space \textbf{C}ausal \textbf{M}ulti-\textbf{T}ask framework, namely DCMT, for post-click conversion prediction. To debias the selection bias in the entire exposure space $\mathcal{D}$, we propose a new counterfactual mechanism with our proposed structure of twin tower, which can effectively improve the prediction accuracy and avoid the over-fitting problem caused by selection bias and data sparsity. Also, we have conducted extensive experiments to demonstrate the superior performance of our proposed DCMT model. In the future, we plan to study the effect of different counterfactual strategies on our DCMT's performance.